\documentclass[showpacs,twocolumn]{revtex4-1}
\usepackage{graphicx}
\usepackage{dcolumn}
\usepackage{amsmath}
\usepackage[latin1]{inputenc}
\usepackage{graphicx}
\usepackage{amssymb}
\usepackage[colorlinks=true, citecolor=blue, urlcolor = blue, linkcolor= red, bookmarks=true]{hyperref}
\usepackage{float}
\usepackage{amsmath}
\usepackage{amsfonts}
\usepackage{dcolumn}
\usepackage{hyperref}
\usepackage{subfigure}
\usepackage{pgfplots}
\usepackage{epstopdf}
\usepackage{booktabs}
\def \beq{\begin{equation}}
\def \eeq{\end{equation}}
\def \bse{\begin{subequations}}
\def \ese{\end{subequations}}
\def \bea{\begin{eqnarray}}
\def \eea{\end{eqnarray}}
\def \bem{\begin{displaymath}}
\def \eem{\end{displaymath}}
\def \bem{\begin{pmatrix}}
\def \eem{\end{pmatrix}}
\def \beb{\begin{bmatrix}}
\def \eeb{\end{bmatrix}}
\def \bc{\begin{center}}
\def \ec{\end{center}}

\def \nn{\nonumber}

\newcommand{\fixme}[1]{{\color{black}#1}}

\makeatletter
\def\btt#1{\texttt{\@backslashchar#1}}
\DeclareRobustCommand\bblash{\btt{\@backslashchar}} \makeatother

\makeatletter
\def\btt#1{\texttt{\@backslashchar#1}}
\DeclareRobustCommand\bblash{\btt{\@backslashchar}} \makeatother

\begin{document}
	
\title{Rotating black strings in de Rham-Gabadadze-Tolley massive gravity}
		
\author{Sushant G. Ghosh} \email{sghosh2@jmi.ac.in, sgghosh@gmail.com }
\affiliation{Centre of Theoretical Physics, Jamia Millia Islamia, New Delhi 110025, India}
\affiliation{Astrophysics and Cosmology Research Unit, School of Mathematical Sciences, University of Kwazulu-Natal, Private Bag 54001, Durban 4000, South Africa}

\author{Rahul Kumar} \email{rahul.phy3@gmail.com }
\affiliation{Centre of Theoretical Physics, Jamia Millia Islamia, New Delhi 110025, India}

\author{Lunchakorn Tannukij} \email{l\_tannukij@hotmail.com}
\affiliation{ Department of Physics, Hanyang University, Seoul 133-891, South Korea}
\affiliation{Theoretical and Computational Physics Group, Theoretical and Computational Science Center(TaCS), Faculty of Science, King Mongkut's University of Technology Thonburi, 126 Pracha Uthit Rd., Bang Mod, Thung Khru, Bangkok 10140, Thailand}

\author{Pitayuth Wongjun} \email{pitbaa@gmail.com}
\affiliation{The institute for fundamental study, Naresuan University, Phitsanulok 65000, Thailand}
\affiliation{Thailand Center of Excellence in Physics, Ministry of Education, Bangkok 10400, Thailand}	
	

\begin{abstract}
One of the solutions of Einstein field equation with cylindrical symmetry is known as black string solution. In this work, the rotating black string solution in de Rham-Gabadadze-Tolley (dRGT) massive gravity is obtained and then called rotating-dRGT black string solution. This solution is a kind of generalized version of rotating anti-de Sitter (AdS)/dS black string solution containing an additional two more terms characterizing the structure of graviton mass. The horizon structures of the black string are explored. The thermodynamical properties of the black string are investigated. We found that it is possible to obtain the Hawking-Page phase transition depending on the additional structure of the graviton mass, while it is not possible for usual rotating-AdS/dS black string. By analyzing the free energy, we also found that the stable rotating black string is bigger than the nonrotating one.   
\end{abstract}

\maketitle

\section{Introduction}
Massive gravity theory is a theory that extends Einstein's general relativity (GR) by adding consistent interaction terms interpreted as a graviton mass. Such a theory can provide the solution for describing our Universe, which is currently expanding with acceleration without introducing a cosmological constant. Massive gravity modifies the gravity by weakening it at the large scale compared to GR, which allows the Universe to accelerate whereas the predictions at small scale are kept to be the same as those in GR. The cost of introducing the mass to the graviton is that it breaks the diffeomorphism invariance which normally resides in GR. The first attempt was done in 1939 by Fierz and Pauli \cite{Fierz:1939ix}; they added the interaction terms in the linearized level of GR and later on it was found that the theory made by Fierz and Pauli suffered from the discontinuity in predictions which were pointed out by van Dam, Veltman, and Zakharov, the so-called van Dam-Veltman-Zakharov (vDVZ) discontinuity \cite{VanNieuwenhuizen:1973fi,vanDam:1970vg,Zakharov:1970cc}. This discontinuity problem invoked further studies on the nonlinear generalization of Fierz-Pauli massive gravity. Boulware and Deser found that such nonlinear generalization can only generate an equation of motion which has higher derivative term yielding a ghost instability in the theory, later called Boulware-Deser (BD) ghost \cite{Boulware:1973my}. In the same time, Vainshtein found that the origin of the vDVZ discontinuity is that the prediction made by the linearized theory cannot be trusted inside some characteristic ``Vainshtein'' radius and he also proposed the mechanism that can be used to recover the prediction made by GR for the nonlinear massive gravity \cite{Vainshtein:1972sx}.

Recently, these main problems of massive gravity could be solved by de Rham \textit{et al.} \cite{deRham:2010ik,deRham:2010kj}. They introduced the massive gravity action which contains a nonlinear interaction term which is free from BD ghost and also admits the Vainshtein mechanism. The de Rham-Gabadadze-Tolley (dRGT) massive gravity is well constructed so that the equations of motion contains no higher derivative term to avoid BD ghost. As a consequence, such construction gives rise to a certain energy scale to the dRGT massive gravity formally known as $\Lambda_3$ scale. This scale can be parametrically expressed as $\Lambda_3=\left(M_{Pl}m^2_g\right)^{1/3}$ which marks the cutoff, or, in other words, strong coupling scale, of this dRGT theory when viewed as an effective theory. The reviews on these topics are in Refs.~\cite{Hinterbichler,deRham:2014zqa}.

Beside the cosmological solutions, there have been various investigations of spherically symmetric solutions \cite{Koyama:2011yg,Koyama:2011xz,Nieuwenhuizen:2011sq,Vegh:2013sk,Tasinato:2013rza}. These solutions allows us to investigate the properties of the local  astronomical objects such as the white dwarfs \cite{EslamPanah:2018evk}, neutron star\fixme{s} \cite{Hendi:2017ibm}, and black hole\fixme{s}  \cite{Berezhiani:2011mt,Brito:2013xaa,Volkov:2013roa,Cai:2012db,Babichev:2014fka,Babichev:2015xha,Hu:2016hpm}. Thermodynamical properties of the black hole are also intensively investigated \cite{Cai:2014znn,Ghosh:2015cva,Adams:2014vza,Xu:2015rfa,Capela:2011mh,Hu:2016mym,Zou:2016sab,Hendi:2017arn,Hendi:2017bys,EslamPanah:2016pgc,Hendi:2016hbe,Hendi:2016uni,Hendi:2016yof,Arraut:2014uza,Arraut:2014iba}. Other properties, such as the superradiant effect \cite{Burikham:2017gdm} and greybody factor \cite{Boonserm:2017qcq}, of the black holes in dRGT massive gravity and  the mass-radius ratio bounds for compact objects \cite{Kareeso:2018xum} are investigated. The modification of the gravity due to the graviton mass as a dark matter is also determined in terms of the rotation curves of galaxies \cite{Panpanich:2018cxo}. Furthermore, the motion of a particle around the spherical object is shown to be affected by the modification of the graviton mass \cite{prepare}.

It is well known that the usual observed astronomical objects do not respect the static and spherical symmetry; they are commonly known as rotating prolate spheroids.
Therefore, this allows investigating the astronomical objects satisfying the cylindrical symmetry. Theoretically, the study of the cylindrical solutions provides the better understanding of the hoop conjecture \cite{Thorne}, which states that horizons form when and only when a mass gets compacted into a region whose circumference is less than $4\pi G M$ in all directions.  By using this conjecture, it is expected that the cylindrical matter will not form a black hole. However, it is shown that the hoop conjecture may be violated when the cosmological constant is included since the cylindrical black holes are shown to exist in the GR with the existence of the cosmological constant \cite{Lemos:1994xp,Lemos:1994fn,Cai:1996eg}. The black hole with cylindrical symmetry is called black string. The charged and rotating black string solutions were consequently found \cite{Lemos:1995cm}. The quasinormal modes \cite{Cardoso:2001vs} and the greybody factor of the black string have been investigated \cite{Ahmed}. 

For the dRGT massive gravity theory, it is found that the spherically symmetric solution can provide a more general solution than the Schwarzschild-de Sitter (dS)/anti-de Sitter (AdS). Therefore, it is possible to obtain the cylindrical solution or black string in the dRGT massive gravity theory \cite{Tannukij:2017jtn}. From this investigation, it is found that the Hawking-Page phase transition \cite{Hawking:1982dh,York:1986it}, a transition from the non-black hole or hot flat space state to a black hole, can be obtained while it is not possible for AdS/dS black string in GR. The quasinormal mode \cite{Ponglertsakul:2018smo} and greybody factor \cite{Boonserm:2019mon} for the dRGT black string solution have been investigated as well. In this work, we investigate the rotating black string solution in dRGT massive gravity theory. The horizon structures in both charged case and the uncharged case of the black string are explored. For asymptotically dS black string, it is possible to obtain two horizons while it is not for the usual dS black string. For asymptotically AdS black string, the maximum number of the horizon is 3 while it is 1 for the usual AdS black string in GR. These modifications come from the existence of the structure of the graviton mass in dRGT massive gravity theory. We discuss this issue in Sec. \ref{Rotating solutions}. We then investigate the thermodynamical properties of the black string in Sec \ref{thermodynamics}. The quantities such as entropy, temperature, mass, heat capacity, and free energy are obtained. The stability of the black string, as well as the possibility to obtain the  Hawking-Page phase transition, is investigated in this section. The results are summarized in Sec. \ref{summary}.

\section{\lowercase{d}RGT massive gravity}\label{model}
We begin by reviewing dRGT massive gravity, which is a well-known nonlinear generalization of a massive gravity and is free of the BD ghost by introducing suitable interaction terms into the Lagrangian. The dRGT massive gravity can be represented as Einstein gravity interacting with the nondynamical field (fiducial or reference metric), and hence its action is the well-known Einstein-Hilbert action plus suitable nonlinear interaction terms as given by \cite{deRham:2010kj}
\begin{eqnarray}\label{action}
 S = \int d^4x \sqrt{-g}\; \frac{1}{2} \left[ R +m_g^2\,\, {\cal U}(g, f)\right],
\end{eqnarray}
where $R$ is the Ricci scalar  and ${\cal U}$ is a potential for the graviton which modifies the gravitational sector with the parameter $m_g$ interpreted as  graviton mass. It is important to note that the form of the fiducial metric $f_{\mu\nu}$ can provide a significant form of the physical metric $g_{\mu\nu}$ \cite{Chullaphan:2015ija,Tannukij:2015wmn,Nakarachinda:2017oyc}. Note that for the following calculations, we adopt the natural unit by which the Newtonian gravitational constant is unity, i.e., $G=1$. The effective potential ${\cal U}$ in four-dimensional spacetime is given by
\begin{eqnarray}\label{potential}
 {\cal U}(g, f) = {\cal U}_2 + \alpha_3{\cal U}_3 +\alpha_4{\cal U}_4 ,
\end{eqnarray}
in which $\alpha_3$ and $\alpha_4$ are dimensionless free parameters of the theory. The dependencies of the terms ${\cal U}_2$, ${\cal U}_3$, and ${\cal U}_4$ on the metric $g$ and scalar fields $\phi^a$ are defined as
\begin{eqnarray}
 {\cal U}_2&\equiv&[{\cal K}]^2-[{\cal K}^2] ,\\
 {\cal U}_3&\equiv&[{\cal K}]^3-3[{\cal K}][{\cal K}^2]+2[{\cal K}^3] ,\\
 {\cal U}_4&\equiv&[{\cal K}]^4-6[{\cal K}]^2[{\cal K}^2]+8[{\cal K}][{\cal
K}^3]+3[{\cal K}^2]^2-6[{\cal K}^4],
\end{eqnarray}
where
\begin{eqnarray}
 {\cal K}^\mu_\nu =
\delta^\mu_\nu-\left(\sqrt{g^{-1}\tilde{f}}\right)^\mu_\nu, \label{K-tensor}
\end{eqnarray}
and the rectangular brackets denote the traces, namely $[{\cal K}]={\cal K}^\mu_\mu$ and $[{\cal K}^n]=({\cal K}^n)^\mu_\mu$. The metric $\tilde{f}_{\mu\nu} = f_{ab} \partial_\mu \phi^a \partial_\nu \phi^b$ can be defined in terms of the reference metric $f_{ab}$ and four scalar fields $\phi^a$ called the St\"uckelberg scalars which are introduced to restore general covariance of the theory. One may recognize the interaction terms as symmetric polynomials of $\cal K$; for a particular order, each of the coefficients of possible combinations is chosen so that these terms do not excite higher derivative terms in the equations of motion of a scalar degree of freedom known as BD ghost.

To proceed further, we choose the unitary gauge $\phi^a=x^\mu\delta^a_\mu$ \cite{Vegh:2013sk}. In this gauge, the tensor $g_{\mu\nu}$ is the observable metric whose linear fluctuations around some certain background describe the five propagating degrees of freedom of the spin-2 massive graviton. Note that since the St\"uckelberg scalars transform according to the coordinate transformation, once the scalars are fixed, for example,  {due to choosing the} unitary gauge, applying a coordinate transformation will break the gauge  {condition} and then introduce additional changes in the St\"uckelberg scalars. Also, we redefine the two parameters $\alpha_3$ and $\alpha_4$ of the graviton potential in Eq. \eqref{potential} by introducing two new parameters $\alpha$ and $\beta$, as follows:
\begin{eqnarray}\label{alphabeta}
 \alpha_3 = \frac{\alpha-1}{3}~,~~\alpha_4 =
\frac{\beta}{4}+\frac{1-\alpha}{12}.
\end{eqnarray}
By varying the action with respect to metric $g_{\mu\nu}$, we obtain the modified Einstein field equations as
\begin{eqnarray}\label{EoM}
 G_{\mu\nu} +m_g^2 X_{\mu\nu} = 0, \label{modEFE}
\end{eqnarray}
where $X_{\mu\nu}$ is the effective energy-momentum tensor obtained by varying the potential term with respect to $g_{\mu\nu}$,
\begin{eqnarray}
 X_{\mu\nu} &=& {\cal K}_ {\mu\nu} -{\cal K}g_ {\mu\nu} -\alpha\left({\cal K}^2_{\mu\nu}-{\cal K}{\cal K}_{\mu\nu} +\frac{{\cal U}_2}{2}g_{\mu\nu}\right) \nonumber \\
 &&+3\beta\left( {\cal K}^3_{\mu\nu} -{\cal K}{\cal K}^2_{\mu\nu} +\frac{{\cal U}_2}{2}{\cal K}_{\mu\nu} - \frac{{\cal U}_3}{6}g_{\mu\nu} \right). \,\,\,\,\,\,\label{effemt}
\end{eqnarray}
In addition to the modified Einstein equations, one can obtain  {a} constraint by using the Bianchi identities as follows:
\begin{eqnarray}\label{BiEoM}
 \nabla^\mu X_{\mu\nu} = 0,
\end{eqnarray}
where $\nabla^\mu$ denotes the covariant derivative, which is compatible with $g_{\mu\nu}$.
Henceforth, we shall use $\alpha$ and
$\beta$ instead of the  parameters $\alpha_3$ and $\alpha_4$.

\section{Rotating solutions} \label{Rotating solutions}
Most of the real astronomical objects are rotating; therefore, it is worthwhile to investigate the rotating solution of the black string in dRGT massive gravity. However, for the black string solution, it is not difficult to obtain the rotating solution from the nonrotating one since both of them still respect the same symmetry. The general line element for static and cylindrically symmetric spacetime in four dimension reads as
\begin{equation}
 ds^2 = -f(r)dt^2 +\frac{dr^2}{f(r)} +r^2 d\Omega^2,\label{NonRot}
\end{equation}
where $d\Omega^2 =d\varphi^2 + \alpha_g^2 dz^2$ is a metric on the two-dimensional (2D) surface and compatible with black string solution \cite{Lemos:1994xp,Cai:1996eg}. We choose the cylindrical coordinates system such that black string has symmetry along $z-$direction, $-\infty< t< +\infty$, $0\leq r<+\infty$, $-\infty<z<+\infty$, and $0\leq \varphi< 2\pi$.  The solution for the equations of motion in Eq.~(\ref{modEFE}) for ansatz in Eq.~(\ref{NonRot}) has been found in Ref.~\cite{Tannukij:2017jtn}; it yield two possible branches of solution
\begin{eqnarray}
f_1(r) &=&-\frac{b}{\alpha_g r}-\frac{m_{g}^2 r^2 \left(1+\alpha +\alpha ^2-3 \beta \right)}{3 (\alpha +3 \beta) }, \label{solutionf1}\\
f_2(r) &=&-\frac{b}{\alpha_g r}+ m_{g}^2\left(r^2 (1+\alpha +\beta )\right.\nonumber\\
&&- \left. h_0 r (1+2 \alpha +3 \beta )+h_0^2 (\alpha +3 \beta )\right), \label{solutionf2}
\end{eqnarray}
with $b$ as an integration constant expressed as $b=4M$, and $M$ is Arnowitt-Deser-Misner mass per unit length along $z$ direction. Out of these two solutions, only $f_2(r)$ is non-trivial because $f_1(r)$ mimics the Lemos's \cite{Lemos:1994xp} black string solution in AdS background in GR with suitable cosmological constant $\Lambda$  \cite{Tannukij:2017jtn},
\begin{equation}
\Lambda\equiv-3\alpha_g^2=\frac{m_{g}^2 \left(1+\alpha +\alpha ^2-3 \beta \right)}{(\alpha +3 \beta) },
\end{equation}
which has been already studied in literature including their rotating and charged counterparts, and thermodynamical properties \cite{Cai:1996eg}. We must notice that this AdS behavior comes as a natural consequence of non-zero graviton mass in dRGT theory. However, second solution in Eq.~(\ref{solutionf2}) is the generalization of Lemos's black string solution and contains the correction terms due to the dRGT massive gravity theory. Nevertheless, various properties of static solution with metric function $f_2(r)$ have been already studied extensively in Ref.~\cite{Tannukij:2017jtn}. Therefore, we will be focusing on the rotating counterpart of the $f_2(r)$ solution only.  For mathematical simplicity, to rewrite $f_2(r)$ in a more compact form, we can redefine the variables and parameters as follow:
\begin{eqnarray}
f_2({r}) &=&-\frac{b}{\alpha_g r}+ \alpha_{m}^2\left(r^2 -c_1 r +c_0\right),\label{solutionf2-new}
\end{eqnarray}
where
\begin{eqnarray}
\alpha_{m}^2 &\equiv & m^2_g \left(1+\alpha +\beta \right),\,\,\,c_1 \equiv \frac{h_0  (1+2 \alpha +3 \beta )}{1+\alpha +\beta},\nonumber\\
&&  c_0 \equiv \frac{h_0^2 (\alpha +3 \beta )}{1+\alpha +\beta}. \label{defineparameter}
\end{eqnarray}
Note that $b$ is an integration constant and supposed to be $4M$ where $M$ is the black string mass density for non-rotating spacetime. 

It is important to emphasis that there exists a scale such that the graviton mass becomes dominated, $r_V^3 = M/(\alpha_g m_g^2)$, known as the Vainshtein radius. This scale can be obtained by comparing the first term and the second term in function $f_2(r)$ from Eq. (\ref{solutionf2-new}). Note that $M$ is mass per unit length so that $M/\alpha_g$ is effectively the mass of the black string in unit of mass.  In the case at which the graviton mass is the same order of the cosmological constant to obtain the acceleration expansion of the Universe nowadays, $m_g^2 \sim \Lambda \sim H^2$, the radius is expressed as $r_V \sim 10^{16} \text{km}$ where we have used $M$ which is order of the solar mass. As a result, if we set the constants as $M \sim \alpha_g \sim m_g \sim 1$, the length scale is order of the Vainshtein radius. Then most of the plots  shown below are respective to this length scale. In choice of the parameters, one can see that the parameter $\alpha_m$ will characterize the strength of the graviton mass. This parameter plays the same role with cosmological constant parameter $\alpha_g$ in Lemos's solution. The parameters $c_0$ and $c_1$ will characterize the structure of the graviton mass as well as characterize how the model differ from one in the Lemos's solution. Conveniently, we will fix one of them and then vary the other. Specifically, we fix $c_1$ and then vary $c_0$. Note that $c_0$ and $c_1$ are dimensionfull and scaled by the Vainshtein radius, $r_{V}$. In order to see the exact value of the thermodynamics quantities, one may introduce the dimensionless parameters as $\bar{c}_0 = c_0/r_V^2$ and $\bar{c}_1 = c_1/r_V$. Note also that, if we continuously  decreases the parameter $\alpha_m$  until $\alpha_m = 0$, the result is not the solution to the Einstein equation anymore. This is not the case for spherical symmetry in which the Schwarzschild solution will be obtained. This is the crucial differences between solution in spherical symmetry and cylindrical symmetry.

The black string mass density will be modified due to the rotating spacetime depending on the angular frequency, as we will see later. Furthermore, the solution of modified Einstein-Maxwell equations in dRGT massive gravity theory for static and cylindrically symmetric spacetime has been studied \cite{Tannukij:2017jtn}, which reads as 
\begin{eqnarray}
f_2^q(r) &=&-\frac{b}{\alpha_g r}+\frac{\gamma^2}{\alpha_g^2 r^2}+ \alpha_m^2\left(r^2 - c_1 r + c_0 \right),\label{solutionfC}
\end{eqnarray}
with vector potential $A_{\mu}=a(r)\delta^{t}_{\mu}$, where $a(r)$ is arbitrary function of radial coordinate $r$. In order to have the consistent solution with Maxwell equations, $a(r)$ can be interpreted as 
\begin{eqnarray}
a(r) =- \frac{\gamma }{\alpha_g r},
\end{eqnarray}
$\gamma$ being an integration constant can be fixed as $\gamma^2=4 q^2$, where $q$ is identified as the linear charge density in $z-$direction.\newline
The nonrotating black string solution can be extended to the rotating one by using the following simple coordinate transformation \cite{Lemos:1994xp,Stachel:1981fg,Bonnor:1980wm}:
\begin{equation}
t=\lambda\tilde{t} - \frac{\omega}{\alpha_g^2}\tilde{\varphi},\qquad 
\varphi= \lambda \tilde{\varphi} - \omega\tilde{t},\label{trans}
\end{equation}
where $\lambda$ and $\omega$ are constant parameters. On using the coordinate transformation (\ref{trans}), the metric of the physical rotating black string spacetime in dRGT massive gravity is given by
\begin{eqnarray}
ds^2 &=& \left( r^2 \omega^2-\lambda^2 f(r) \right) d\tilde{t}^2 + \frac{dr^2}{f(r)}+\frac{2\lambda \omega}{\alpha^2_g} \left(f(r) - \alpha_g^2 r^2\right)\nonumber\\
&&  d\tilde{t}d\tilde{\varphi}  +\left( r^2 \lambda^2 - \frac{\omega^2}{\alpha^4_g}f(r)\right) d\tilde{\varphi}^2 + r^2 \alpha^2_g dz^2,
\label{bsmetricrotate}\end{eqnarray} 
where $f(r)$ is given by Eqs. (\ref{solutionf2-new}) and (\ref{solutionfC}), respectively, for the charged  and uncharged cases. In the no-rotation ($\omega=0$), the above line element reverts back to the static and cylindrically symmetric black string spacetime \cite{Tannukij:2017jtn}.  On the other hand, the Lemos's \cite{Lemos:1994xp} rotating black string solution in GR can be obtained as a special case of (\ref{bsmetricrotate}) for $c_1=c_0=0$, and $\alpha_g=\alpha_m$.  Likewise, to nonrotating black string, the rotating black string also approach the AdS spacetime for asymptotically large $r$, though for large $z$ with fixed $r$ it do\fixme{es} not approaches the AdS spacetime. The line element Eq. (\ref{bsmetricrotate}) describes a black string rotating only along $\varphi$ direction. For a closed black string with compacted $z$ coordinate $0\leq \alpha_g z<2\pi$ ($S^1\times S^1$ torus topology), we can have rotation along $z$ direction too. This will lead to a black toroid rotating in two orthogonal directions $\varphi$ and $z$. Though it will merely lead to any interesting phenomenon, as we can always make a coordinate transformation to eventually get the black string rotating only along $\varphi$ direction described by Eq. (\ref{bsmetricrotate}) \cite{Lemos:1994xp}. The electromagnetic field also get transformed under the coordinate transformation in Eq.(\ref{trans}) as 
\begin{eqnarray}
A_\mu=(a(r) \lambda ,0,-a(r)\omega /\alpha_g^2,0).
\end{eqnarray}
Since the form of the physical metric depends on the form of the fiducial (reference) metric, the fiducial metric is supposed to get modified in the same way as the physical metric to preserve the equation of motion. According to this coordinate transformation, the fiducial metric can be written as
\begin{eqnarray}
f_{\mu\nu} &=& \left(
\begin{array}{cccc}
 h^2 \omega ^2 & 0 & -h^2 \lambda  \omega  & 0 \\
 0 & 0 & 0 & 0 \\
 -h^2 \lambda  \omega  & 0 & h^2 \lambda ^2 & 0 \\
 0 & 0 & 0 & \alpha_g ^2 h^2
\end{array}
\right) .
\end{eqnarray}
One can check that this form of the fiducial metric still provides the solution in Eq. (\ref{bsmetricrotate}) of the modified Einstein equation in (\ref{modEFE}).

Naively, this solution may not be considered as the rotating solution since the solution is obtained by using the coordinate transformation. However, by applying the inverted coordinate transformation, the original nonrotating one cannot be obtained as we expect. It is worth mentioning that the periodic nature of $\varphi$ prevents $(t,\varphi)\to (\tilde{t},\tilde{\varphi})$ to be a proper global coordinate transformation in the entire manifold, rather this can be done only locally \cite{Lemos:1994xp,Stachel:1981fg,Bonnor:1980wm}.  For spacetime associated with solution in the cylindrical symmetry,  there exists a closed curve which cannot be continuously shrunk to a point, a closed curve warps around the $z$ axis. This kind of spacetime is said to be not simply connected. Since the spacetime is not simply connected, the coordinate transformation can be done locally but not for the entire manifold. Then the coordinate transformation cannot be done globally for such kind of spacetime. Therefore, the spacetime obtained  by using the above coordinate transformation is distinct from the original one.

It is well known that Schwarzschild solution is a unique solution of Einstein field equations for spherical symmetry. At the event horizon, the timelike Killing vector is null. This allows us to define the surface gravity and then temperature of the black hole. Similarly, for axial symmetry and stationary spacetime, there exist two Killing vectors: timelike vector $\eta^\mu_{(t)}$ corresponding to generator of the time translation and the other $\eta^\mu_{(\varphi)}$ corresponding to the generator of the rotation. The unique solution for this kind of symmetry is Kerr solution. It is found that, in Kerr black hole, a Killing vector which form by a linear combination of two Killing vectors with a special coefficient $\xi^\mu = \eta^\mu_{(t)}+\Omega_H \eta^\mu_{(\varphi)}$ is null at the event horizon where $\Omega_H$ is the angular velocity of a particle at the event horizon. Therefore, the surface gravity of the Kerr black hole can be defined via this Killing vector. Actually, the existence of the Killing vector $\xi^\mu$ implies the existence of the rotating solution. This situation is similar to our case. We found the Killing vector which is a linear combination of the two Killing vectors. This Killing vector is null at the event horizon. As a result, this implies the existence of the stationary spacetime or rotating solution. This allows us to calculate the temperature of the black string as well as the quantities corresponding to the rotation such as the angular momentum of the black string as shown later in Eq. (\ref{angular1}).

\begin{figure*}[!ht]
\begin{tabular}{c c}
\includegraphics[scale=0.7]{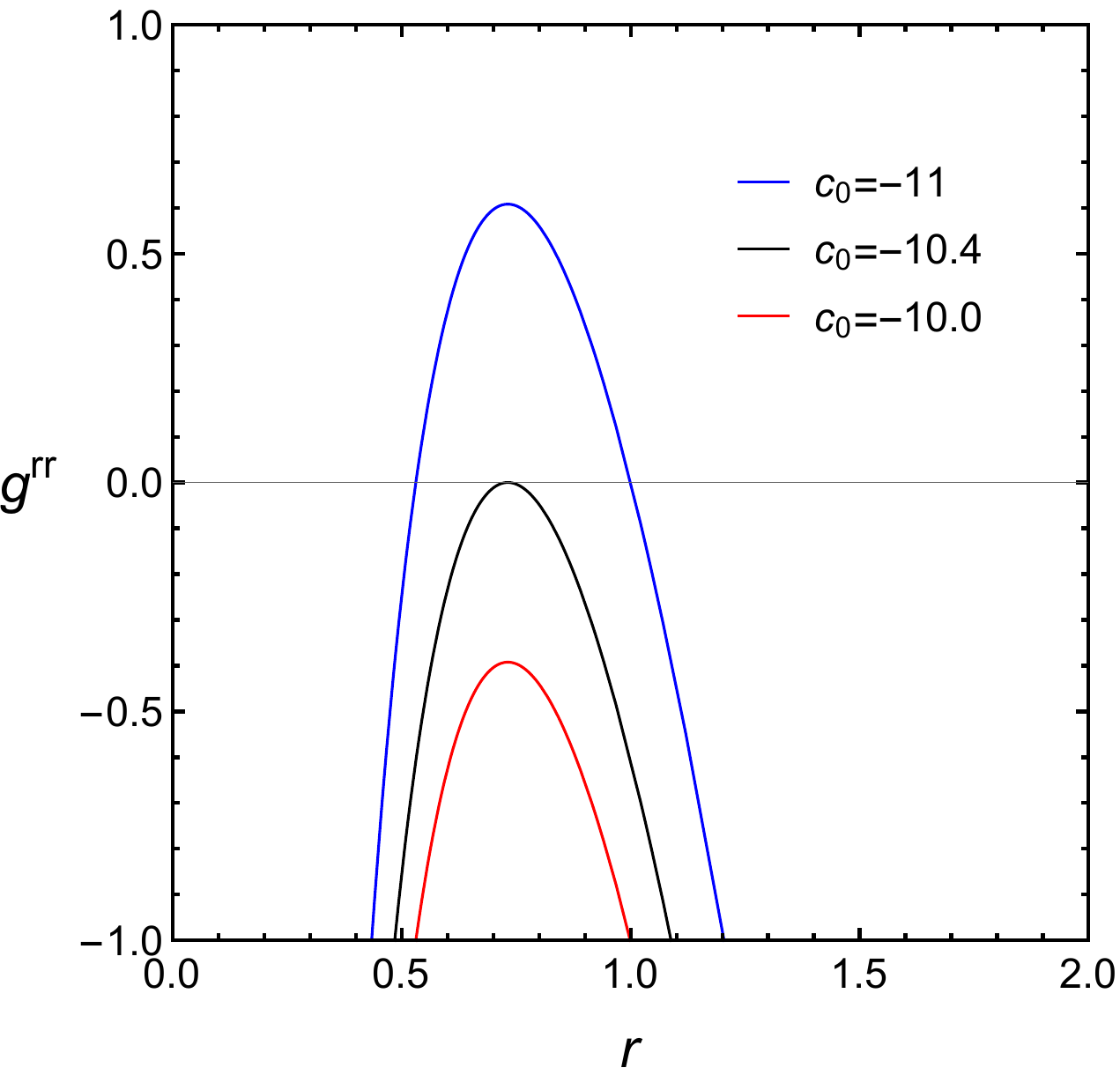}\quad
\includegraphics[scale=0.67]{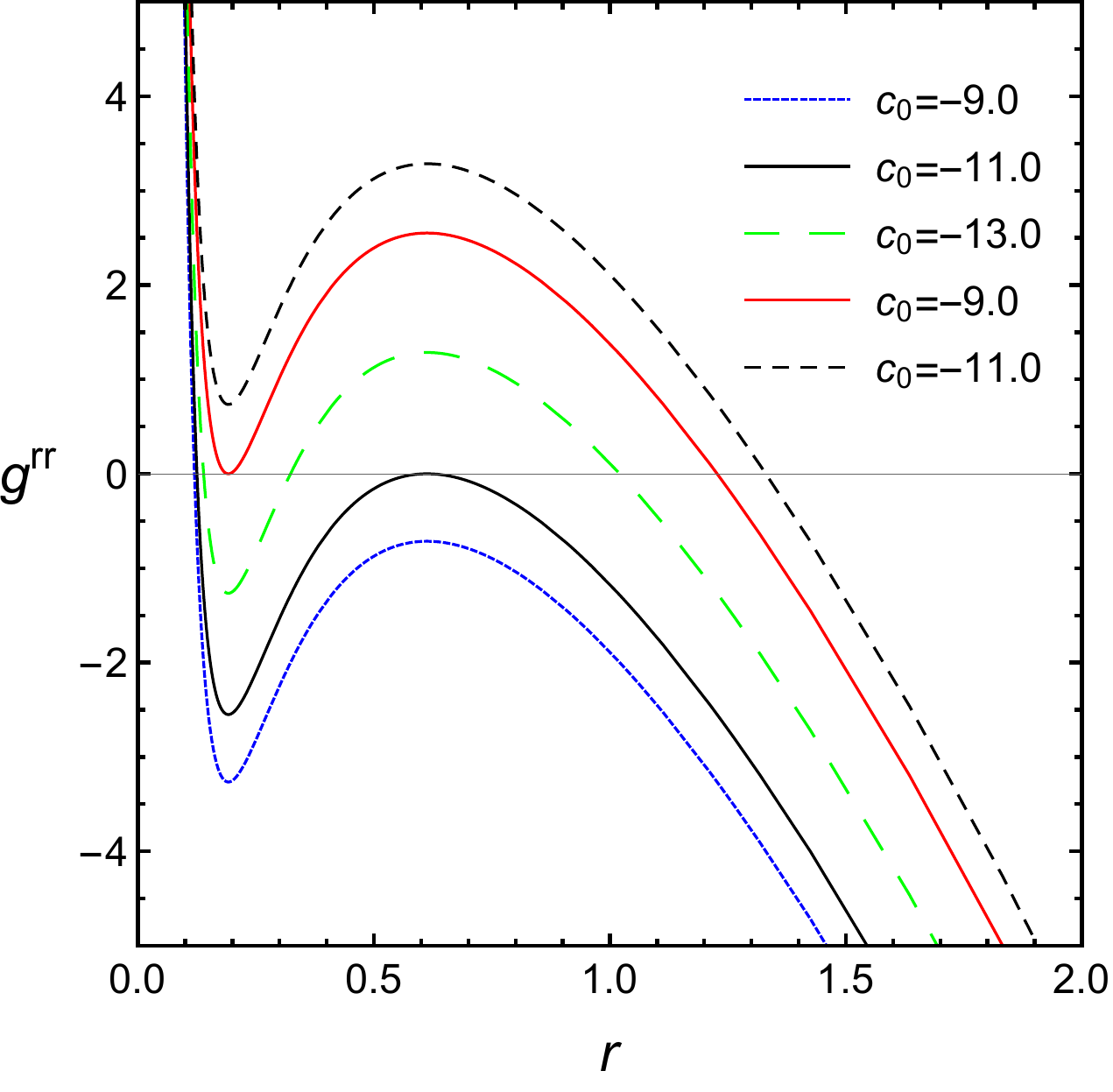}
\end{tabular}
 \caption{Plot of $g^{rr}$ vs $r$ for asymptotically dS rotating black string in dRGT massive gravity for particular values of parameters $b=4, \alpha_g=1, \alpha_m^2 =-1, c_1=-6$. The left panel is the uncharged case and the right is the charged case with $q=0.3$.}\label{EH-dS}
\end{figure*}

\begin{figure*}[!ht]
\begin{tabular}{c c}
\includegraphics[scale=0.67]{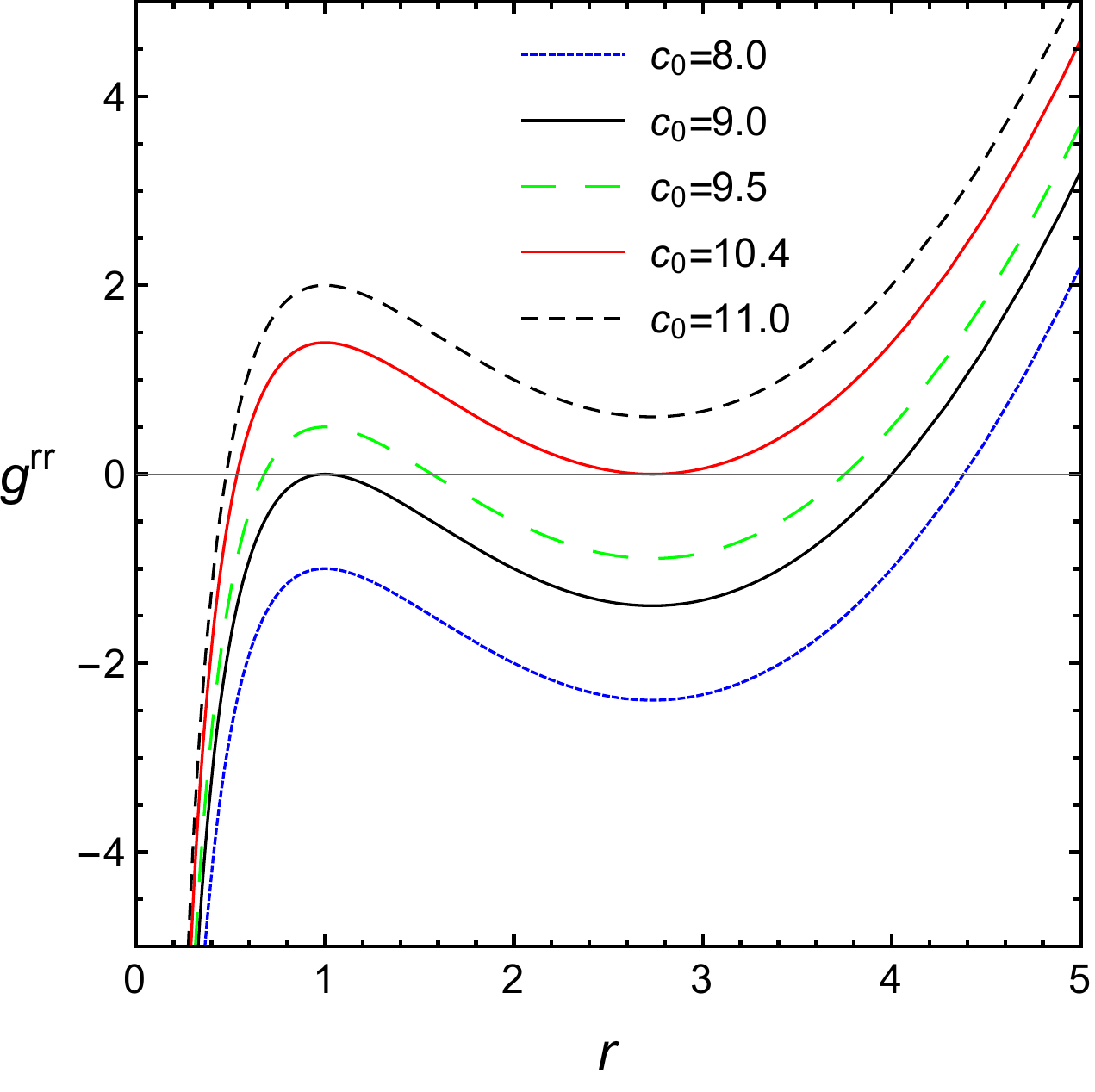}&
\includegraphics[scale=0.67]{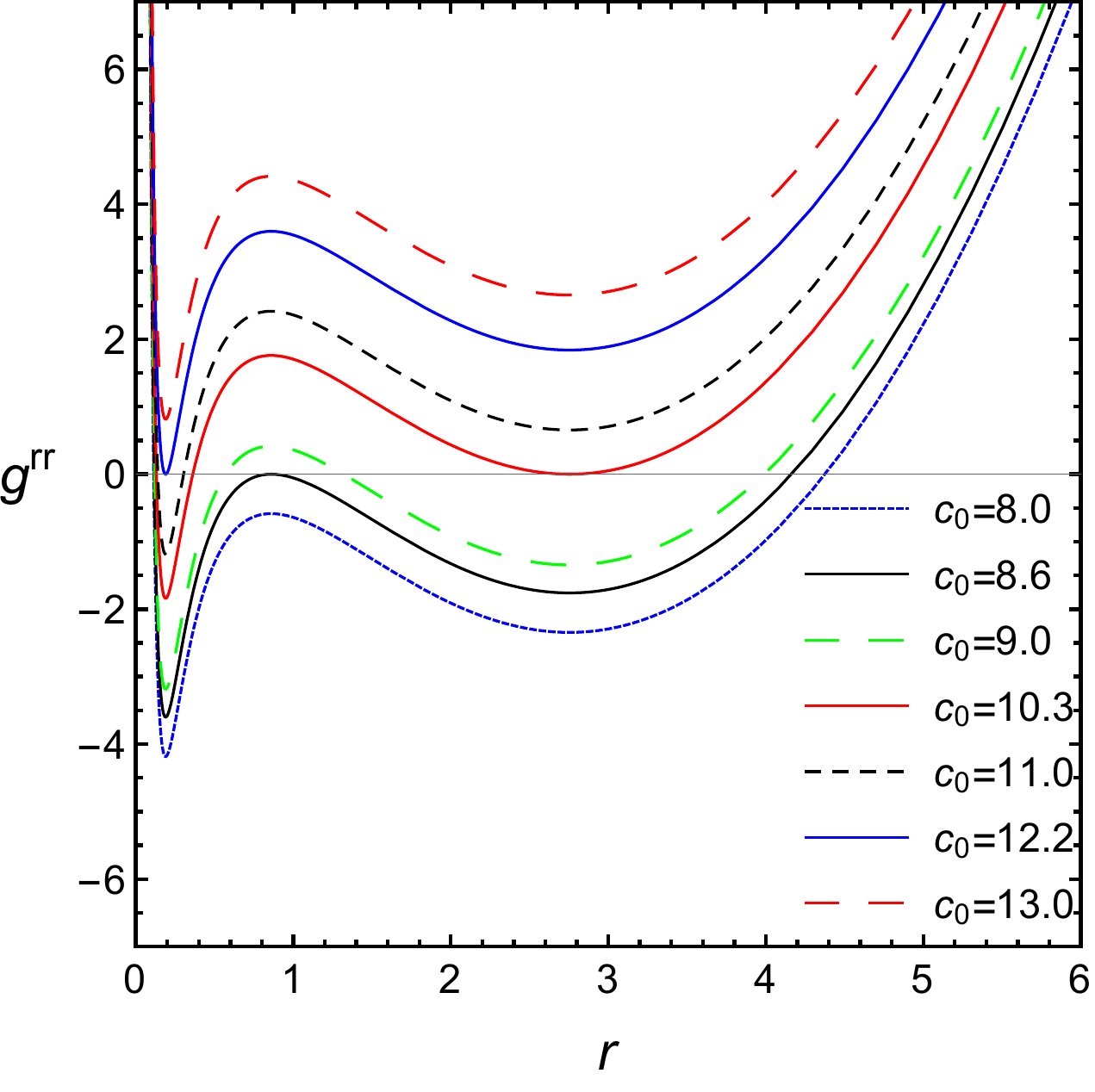}
\end{tabular}
 \caption{Plot of $g^{rr}$ vs $r$ for asymptotically AdS rotating black string in dRGT massive gravity for particular values of parameters $b=4, \alpha_g=1, \alpha_m^2 =1, c_1=6$. The left panel is the uncharged case and the right is the charged case with $q=0.3$.}\label{EH-AdS}
\end{figure*}

The rotating black string metric has coordinate singularity at $g^{rr}=0\Rightarrow f(r)=0$, whose solutions determined the radial coordinates of horizons, viz., $f_2(r)=0$ for uncharged rotating black string and $f_2^q(r)=0$ for charged one. Clearly, the number of horizons and their positions have an explicit dependency upon the parameters $b, \alpha_g, \alpha_m,$ $ q, c_1, c_0$, and crucially on the sign of parameter $\alpha_m^2$. For $\alpha_m^2 < 0$, the spacetime is asymptotically dS and then the number of horizons is upto two for the uncharged case and up to three for the charged case as shown in  Fig. \ref{EH-dS}. Whereas for the asymptotically AdS case, $\alpha_m^2 > 0$, the number of horizons is up to three for the uncharged case and up to four for charged case (cf. Fig.~\ref{EH-AdS}). In Figs. \ref{EH-dS} and \ref{EH-AdS}, we plotted $f = g^{rr}$ vs $r$ for uncharged and charged black string $(q=0.3)$ in left and right panel, respectively. It is also shown that there exist critical values where two horizons merge together corresponding to the extremal case.
Let us point the important issue for the rotating black string in dRGT massive gravity. As we have mentioned, the crucial difference in this solution to the usual black string is the existence of $c_0$ and $c_1$ terms. Without these two terms, it is not possible to have the horizons in uncharged and asymptotically dS case, though the structure of the graviton mass allows the existence of the horizon. However, for the asymptotically AdS case, one horizon can always be found even without these two terms. This will significantly affect the thermodynamics behavior as we will see later. Furthermore, for the charged case, it provides more horizons and then the thermodynamics is significantly changed. 

\begin{figure*}[!ht]
\begin{tabular}{c c}
\includegraphics[scale=0.71]{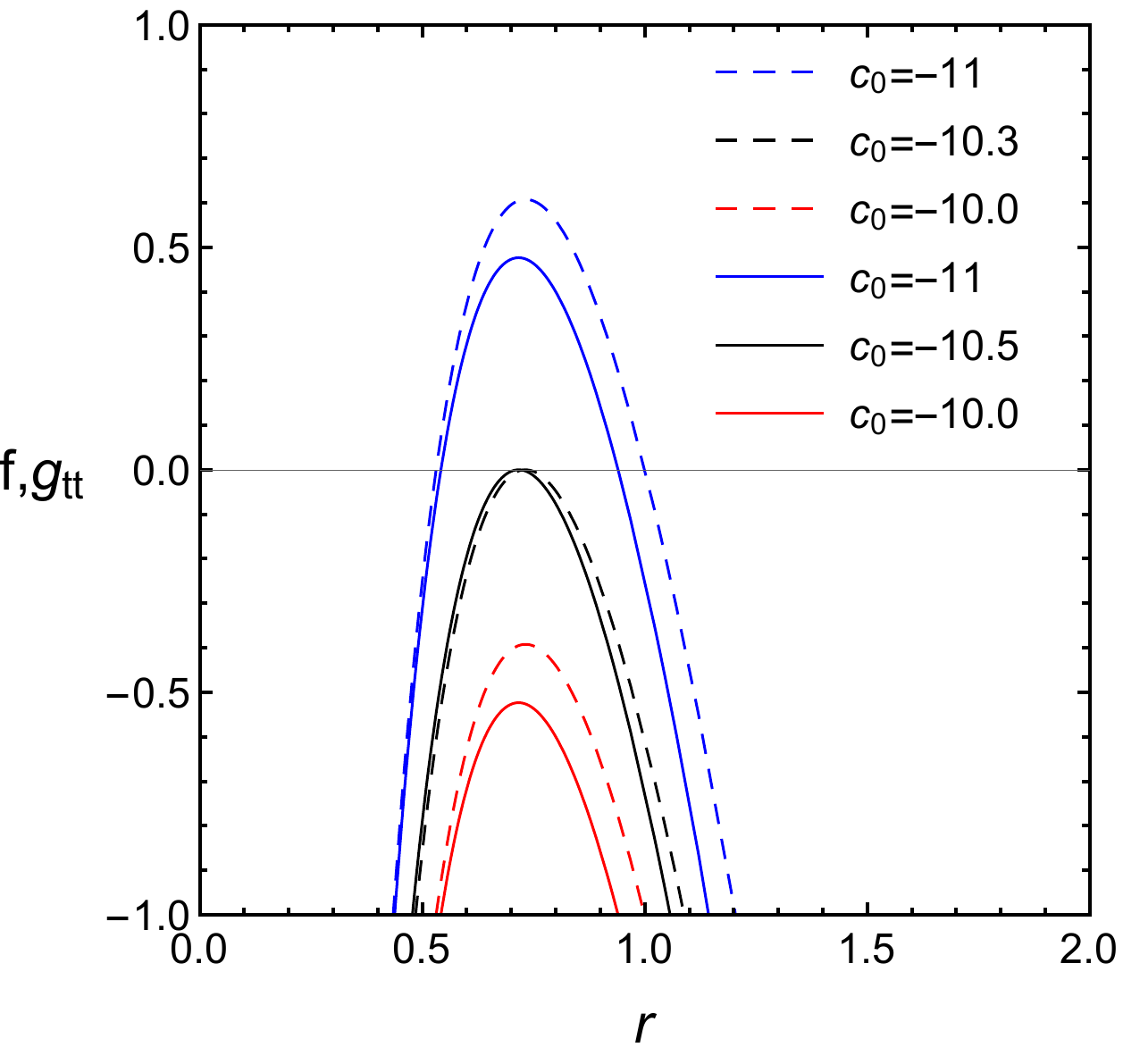}\quad
\includegraphics[scale=0.65]{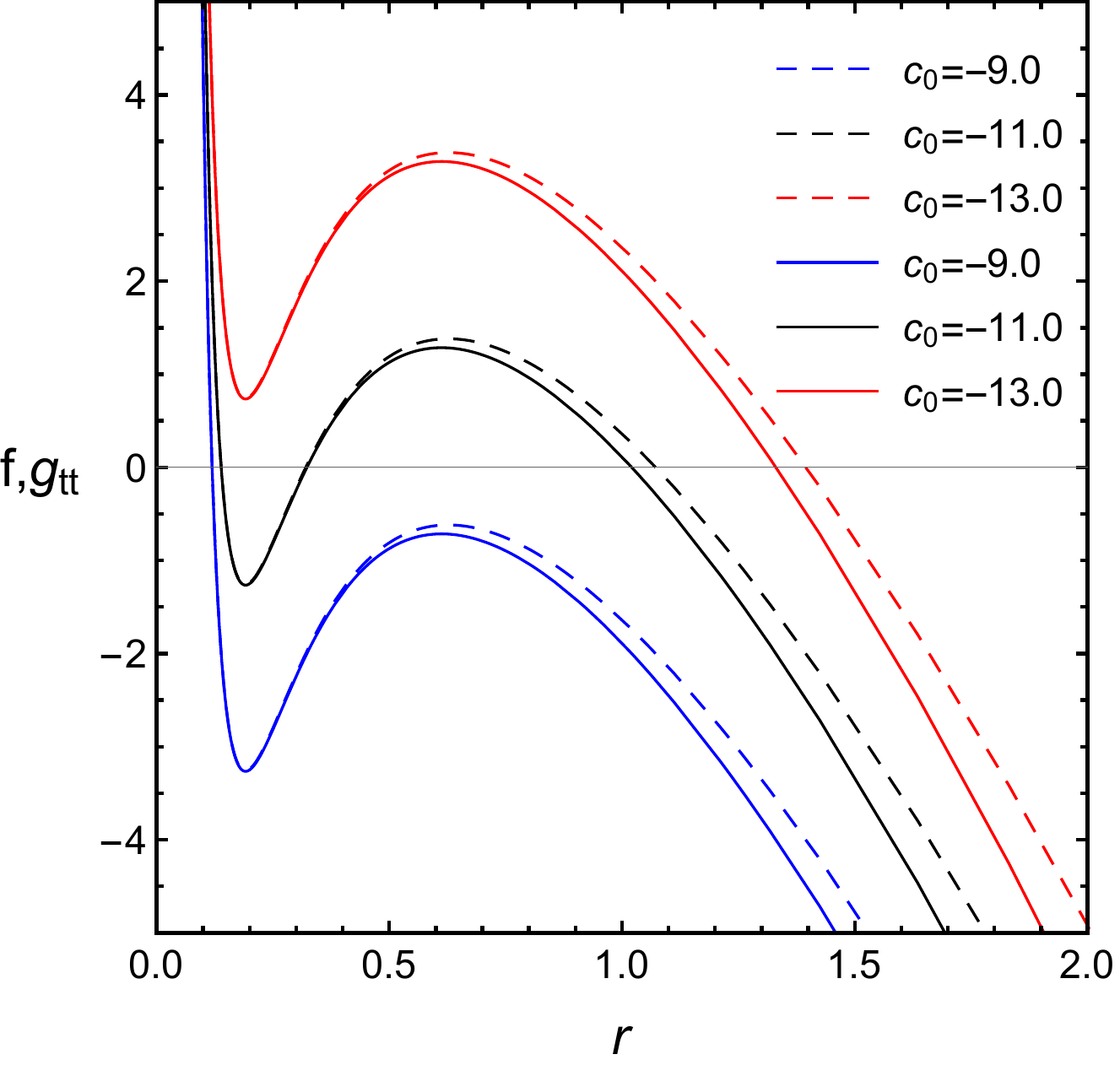}
\end{tabular}
 \caption{Plot of $g_{tt}$ and $g^{rr}$ vs $r$ for asymptotically dS rotating black string in dRGT massive gravity for particular values of parameters $b=4, \alpha_g=1, \lambda=1, \alpha_m^2 =-1, c_1=-6, \omega = 0.5$. The solid lines represent $g^{rr} = f$ corresponding to horizons with $f=0$ while dashed lines represents $g_{tt}$ corresponding to SLS with $g_{tt}=0$. The left panel is the uncharged case and the right is the charged case with $q=0.3$.}\label{SLS-dS}
\end{figure*}

\begin{figure*}[!ht]
\begin{tabular}{c c}
\includegraphics[scale=0.6]{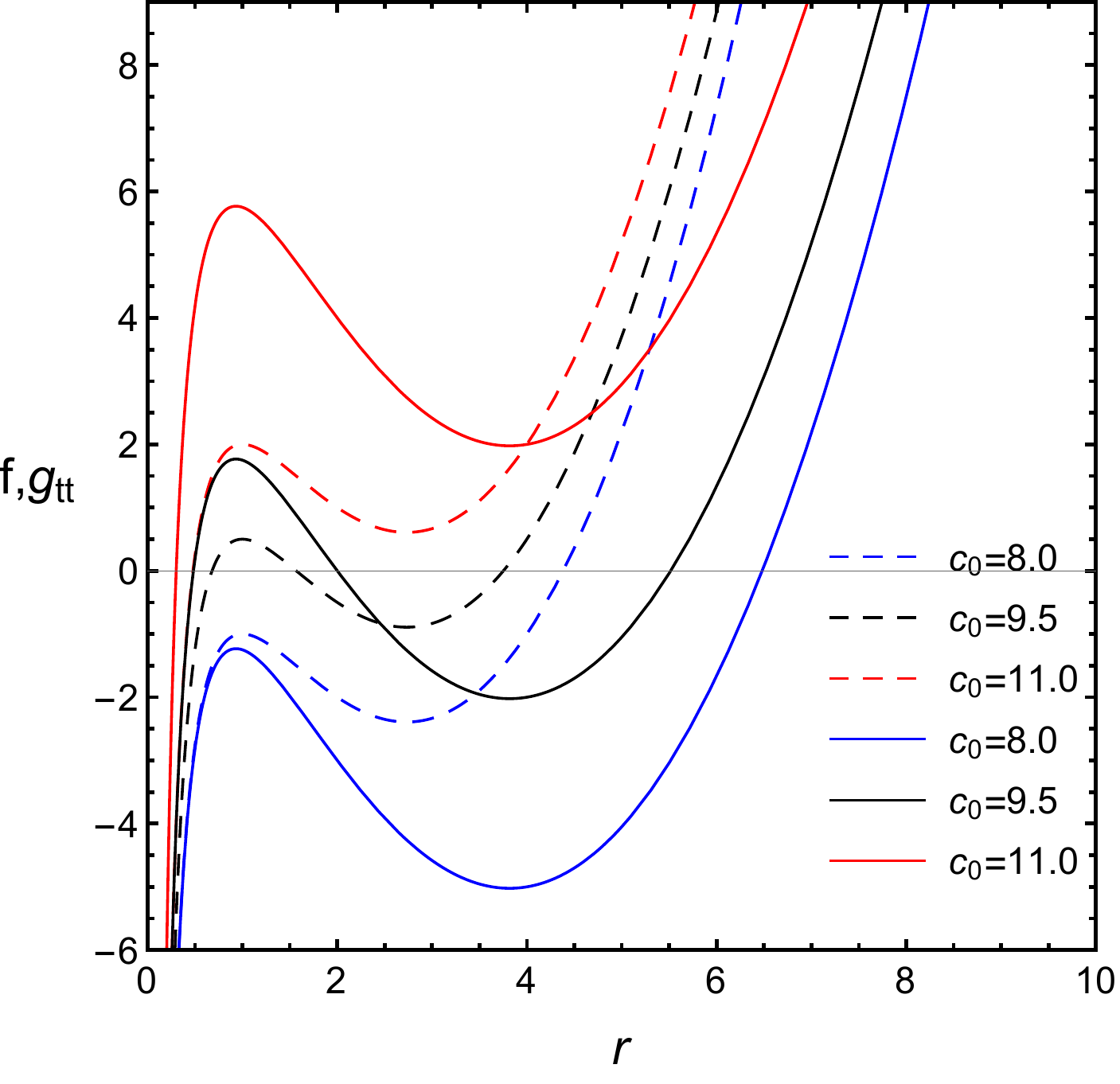}\quad
\includegraphics[scale=0.6]{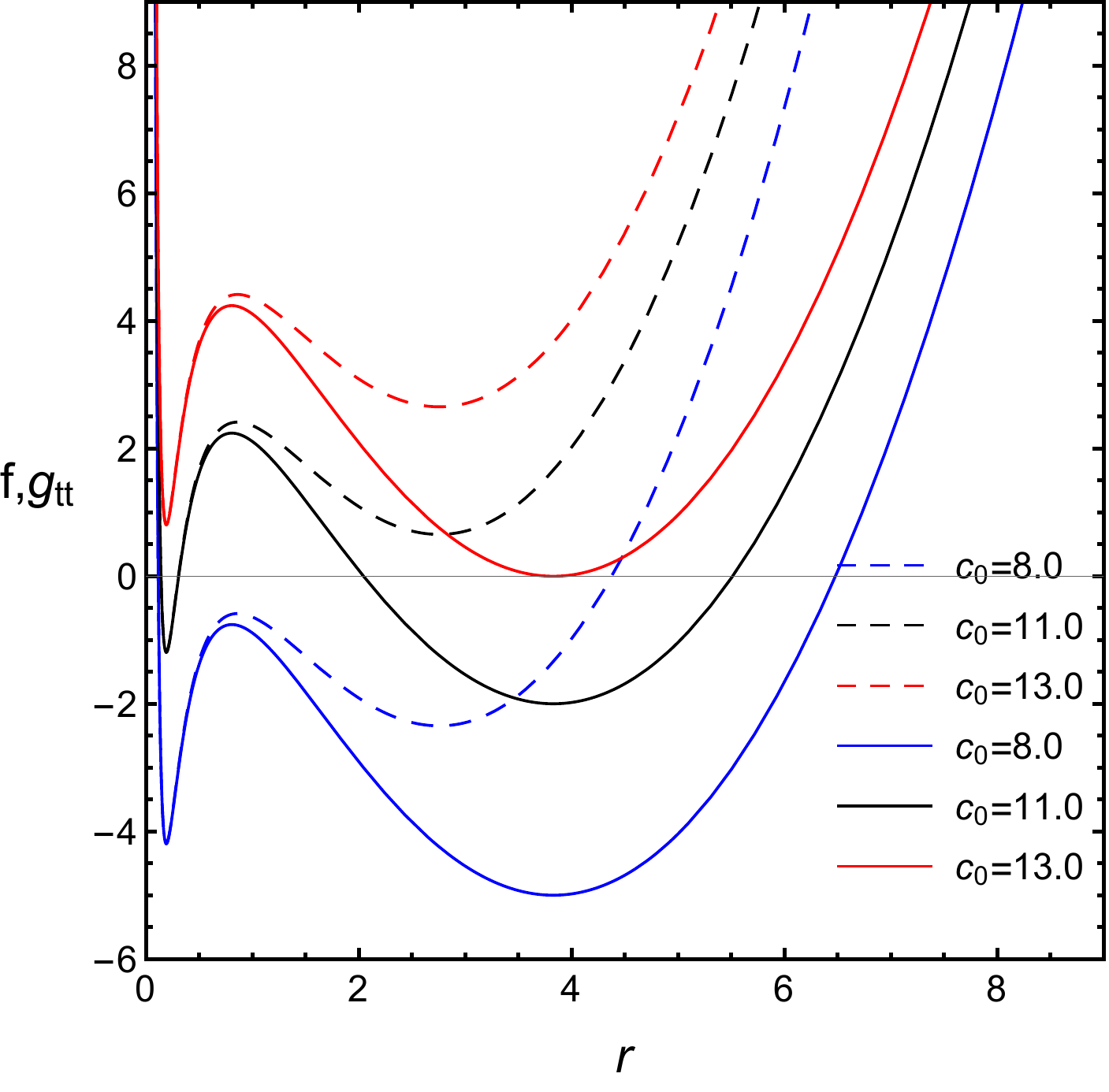}
\end{tabular}
 \caption{Plot of $g_{tt}$ and $g^{rr}$ vs $r$ for asymptotically dS rotating black string in dRGT massive gravity for particular values of parameters $b=4, \alpha_g=1, \lambda=1, \alpha_m^2 =1, c_1=6, \omega = 0.2$. The solid lines represent $g^{rr} = f$ corresponding to horizons with $f=0$, while dashed lines represents $g_{tt}$ corresponding to SLS with $g_{tt}=0$. The left panel is the uncharged case and the right is the charged case with $q=0.3$.}\label{SLS-AdS}
\end{figure*}

This is evident from Eq.~(\ref{bsmetricrotate}), that rotating black string also has another physical relevant surface called "static limit surface" (SLS) governed by the solution of $g_{tt}=0 \Rightarrow\lambda^2 f(r)-r^2 \omega^2=0$, which coincides with the event horizon in the nonrotating limit ($\omega=0$). As a result, we have two more parameters in order to find the behavior of the SLS. However, the structure of the surface does not significantly change since the additional terms are proportional to $r^2$ which already exist in $f(r)$. Therefore, the number of SLS is the same as those for horizons and we have shown it explicitly. Specifically, we plot how the horizon changes when the term $r^2 \omega^2$ is added as shown in Fig \ref{SLS-dS}. Note that the other structures, for example, the existence of the extremal case, will not be significantly changed. For asymptotically AdS case, the value of $\omega$ significantly changes the structure of the SLS surface since the $\omega^2r^2$ term in  $\lambda^2\,f -\omega^2 r^2 $ will cancel the contribution in graviton mass in contrast to the dS case, which it will support. As a result, we use $\omega = 0.2$ instead of $\omega = 0.5$ for dS case as shown in Fig. \ref{SLS-AdS}. 

 
The charge and mass linear densities come naturally as the integration constants in terms of $\gamma$ and $b$, respectively. Though the integration constants in both Eq. (\ref{solutionf2}) and Eq. (\ref{solutionfC}) may not necessarily be the same as in the stationary solution. 
In order to find the proper constants, we can find the relation of the constants to the physical mass, angular momentum, and charge of the black string. The metric described in Eq. (\ref{bsmetricrotate}) has an infinite extension along $z$ direction; therefore, it looks obvious that for a far distant observer ($r\rightarrow \infty$), the total mass and total charge would be infinite. The physical quantities, in this case, are mass and charge linear densities, which are finite. In order to estimate these quantities, we can use the Hamiltonian formalism as suggested by Brown and York \cite{Brown:1992br}.
We can redefine the metric into the canonical form as follows:
\begin{eqnarray}
ds^2 &=& -N^2_0 d\tilde{t}^2 + R^2 (N_\varphi d\tilde{t} \nonumber\\
&& + d\tilde{\varphi})^2 + \bar{f}^{-2} dR^2+ r^2 \alpha_g^2 dz^2, \label{AMD form}
\end{eqnarray}
where
\begin{eqnarray}
N^2_0 &=&\frac{r^2}{R^2} \Delta^4  f(r),\,\,\Delta^2 = \lambda^2 - \frac{\omega^2}{\alpha^2_g},\,\,\bar{f}^2 = \left(\frac{dR}{dr}\right)^2 f(r),\nonumber\\
R^2 &=& \lambda^2 r^2 -\frac{\omega^2}{\alpha^4_g}f(r),\,\, N_\varphi = \frac{\lambda \omega}{\alpha^2_g R^2}\left(f(r)- \alpha^2_g r^2 \right),\nonumber\\
\Delta_m^2 &=& \lambda^2 - \frac{\omega^2 \alpha_m^2}{\alpha^4_g}\left( 1 - \frac{c_1}{r} +\frac{c_0}{r^2}\right).
\end{eqnarray}
Here, $N_0$ and $N_{\varphi}$ are, respectively, called the lapse and shift functions.
In order to estimate the physical quantities, we consider a $\tilde{t}=$ constant hypersurface $\Sigma$, which foliate the four-dimensional manifold $\cal{M}$, and described by the metric $h_{ij}$ and future pointing unit normal $u^{\mu}$. An element of three-boundary of $\cal{M}$, $^3\mathcal{B}$ is a timelike three-surface generated by metric $\gamma_{ij}$, and outward unit normal vector $n^{\mu}$. Let $\sigma$ is the determinant of the metric $\sigma_{ij}$ evaluated on two-boundary $^2\mathcal{B}$ of hypersurface $\Sigma$ with constraints $dR=0$ and $d\tilde{t}=0$. With such construction, we can determine the proper surface energy density by the projection of stress tensor defined on three-boundary $^3\cal{B}$ to the normal of two-boundary $^2\cal{B}$ \cite{Brown:1992br}. Furthermore, the conserved charges can be defined in terms of the Killing vectors on the boundary of the surface and surface stress tensor, which are equal to the value of Hamiltonian required to generate the diffeomorphism along the Killing vectors \cite{Brown:1992br}. In the black string spacetime, $^2\cal{B}$ resemble the surface of an infinite cylinder. In order to make the region finite, we consider that $^2\cal{B}$ is bounded also between $z=z_1$ and $z=z_2$. This is expected that the two obvious Killing vectors $\xi^{\mu}_{(t)}$ and $\xi^{\mu}_{(\phi)}$ of metric (\ref{bsmetricrotate}) corresponding to the time translation and rotation invariance, respectively, will entail the existence of two conserved quantities, which can be identified as energy (or mass) and angular momentum. The global conserved charge $Q_\xi$ associated with metric (\ref{bsmetricrotate}) can be written as \cite{Brown:1992br}
\begin{eqnarray}
Q_\xi = \int d^2x \sqrt{\sigma} (\epsilon u^\mu + j^\mu) \xi_\mu,\label{conserved}
\end{eqnarray}
where $\epsilon$ and $j^\mu = (0,j^i)$ are, respectively, the energy and momentum surface densities on the two-surface $^2\mathcal{B}$, which are defined as
\begin{eqnarray}
\epsilon &=& \frac{k}{8\pi } = \frac{\nabla_i n^i}{8\pi },\\
j^i &=& \frac{h^i_j n_k \Pi^{jk}}{\sqrt{h} 16 \pi } = \frac{h^i_j n_k (K h^{jk} - K^{jk})}{8 \pi },
\end{eqnarray}
where $K_{\mu\nu}$ and $k_{ab}$ are extrinsic curvature of hypersurface $\Sigma$ embedded in $\cal{M}$ and of $^2\cal{B}$ embedded in $\Sigma$, respectively, whereas $K$ and $k$ are their respective traces. $n^i$ is a spacelike normal vector to the two-surface $^2\mathcal{B}$ on the three-space, and $\Pi^{ij}$ is the conjugate momentum in the three-space $\Sigma$. In Eq.~(\ref{conserved}), $\epsilon$ and $j^{\mu}$ are defined such that the conserved charge $Q_{\xi}$ does not have any contribution from the background spacetime \cite{Lemos:1994xp}. Following \cite{Lemos:1994xp}, the mass and angular momentum linear densities of black string spacetime at radial infinity ($R\to\infty$) can be written in terms of the constant parameters as follows:
\begin{eqnarray}
M &=& \frac{b}{4}\left( \lambda^2 + \frac{\omega^2 \alpha_{m}^2}{2 \alpha_g^4}\right)\frac{\Delta^2 }{\Delta_{0}^2 },\label{mass1}\\
J &=& \frac{3\lambda \omega b }{8  \alpha_g^2},\label{angular1}
\end{eqnarray}
where $\Delta_{0}^2\equiv\Delta_{m}^2(r\to \infty) = \lambda^2 - \omega^2\alpha_m^2 / \alpha_g^4$. Again, the results can be reduced to those for the black string solution in GR by setting $\alpha_m = \alpha_g$ and $\Delta_{0} = \Delta$.
In the similar fashion, the electric charge of segment $\Delta z$ contributed from the vector potential $A_{\mu}$ can be written as
\begin{eqnarray}
Q &=& \frac{1 }{4\pi}\int d^2x \sqrt{\sigma} \frac{n_i \varepsilon^i}{\sqrt{h}},\nonumber\\
 \varepsilon^R &=& \frac{\alpha_g r R}{f(r) N_0}\left(N_\varphi \partial_R A_\varphi  - \partial_R A_t\right).
\end{eqnarray}
Using the metric form in Eq. (\ref{AMD form}), the electric charge density of black string is computed and expressed in terms of integration constant $\gamma$,
\begin{eqnarray}
q = \frac{Q}{\Delta z} = \frac{\lambda \gamma }{2}\left(1 + \frac{2 \omega^2 (\alpha^2_g - \alpha_m^2)}{\Delta^2 \alpha^4_g}\right).\label{charge}
\end{eqnarray}
From Eq.~(\ref{charge}), one can see that the electric charge reduce to that for nonrotating black string as $\omega=0$ or even $\alpha_m=\alpha_g$.
\section{Thermodynamics of rotating black string} \label{thermodynamics}
Now we are ready to extract some thermodynamical properties of the rotating black string in dRGT massive gravity. First, let us find the black string  mass density obtained by solving $f(r) = 0$ and using Eq. (\ref{mass1}). As a result, the black string mass density can be expressed as
\begin{eqnarray}
M &=& \frac{ \Delta^2\alpha_g r_+}{4\Delta_{0}^2} \left(\lambda^2 +\frac{\omega^2\alpha_m^2}{2 \alpha_g^4}\right) \nonumber\\
&&\times\left(\frac{\gamma^2}{\alpha^2_g r^2_+} +\alpha_m^2 (r_+^2 - c_1 r_+ + c_0)\right).
\end{eqnarray}
For asymptotically dS black string, $\alpha_m^2 < 0$, we have to impose the conditions $2\alpha_g^2/\alpha_m^2 < \omega^2 /(\lambda \alpha_g)^2 < 1$ and $c_0> c_1^2/4$ to obtain the positive definite of the black string mass. However, these conditions do not allow the existence of the horizons. Therefore, we will not consider the thermodynamics properties of the asymptotic dS black string. For asymptotic AdS black string, the conditions to obtain the positive definite of the black string mass can be expressed as $ \omega^2 /(\lambda \alpha_g)^2 < 1, \omega^2 /(\lambda \alpha_g)^2 < \alpha_g^2/\alpha_m^2 $ and $c_0 > c_1^2/4$. These conditions satisfy the existence of the horizons and then allow us to properly investigate the thermodynamical properties of the black string. Note that it is sufficient to restrict our consideration to the case where the angular frequency is sufficiently small so that the conditions are safely satisfied. Note also that by restricting to this consideration, there exists only one horizon so that the temperature of the black string can be uniquely defined.

One of the important thermodynamics quantity is temperature, which can be defined in terms of the surface gravity $\kappa$,
\begin{align}
\xi_\mu \kappa = -\frac{1}{2} \nabla_\mu (\xi_\nu\xi^\nu), \label{def-sur}
\end{align}
where Killing vector $\xi_\mu$ (timelike outside the event horizon) is the generator of event horizon. $\xi^\mu$ is defined in terms of the generator of time translational ($\eta^{\mu}_{(t)}=\delta^{\mu}_t$) and rotational ($\eta^{\mu}_{(\varphi)}=\delta^{\mu}_{\varphi}$) isometries of metric (\ref{bsmetricrotate}),
\begin{equation}
\xi^\mu=\eta^{\mu}_{(t)}+\Omega_{H}\eta^{\mu}_{(\varphi)},
\end{equation}
such that $\xi^\mu$ is orthogonal and null at the horizon, i.e.,
\begin{align}
\xi^\mu\xi_\mu\Big|_{r=\,r_+}=0,\,\,\,\,\Omega_H = -\frac{g_{t\varphi}}{g_{\varphi\varphi} } \Big|_{r=\,r_+}= \frac{\omega}{\lambda}.\label{time-like-Killing}
\end{align}
By contracting $\xi^\mu \kappa$ to Eq. (\ref{def-sur}), the surface gravity can be expressed as follows $\kappa = \frac{1}{2} \frac{\sqrt{g^{RR}}}{\Phi} \partial_{R} \Phi^2\Big|_{r=\,r_+}=\frac{1}{2} \frac{\Delta^2}{\sqrt{g_{\varphi\varphi}}}r_+ f'(r_+)$, where $\Phi^2 = -\xi_\mu\xi^\mu.$ Note that one can find the surface gravity by directly using Eq. (\ref{def-sur}). However, we have to keep in mind that the Killing vector is null at the horizon and then it can be written as $\xi_\mu \propto \partial_\mu R$. The formulation above can be used to calculate the surface gravity in other solution, i.e., Kerr and Schwarzschild solutions. By using $f(r)$, we can calculate the corresponding Hawking temperature from surface gravity,
\begin{eqnarray}
T_+ &=& \frac{\Delta^2}{4 \pi r_+\lambda}\left(\alpha_m^2 (3r_+^2 -2 c_1 r_+ + c_0)- \frac{\gamma^2}{\alpha^2_g r^2_+} \right).
\end{eqnarray}
Note that in the special setting $c_1=c_0 = 0,\; \alpha_g=\alpha_m$, this temperature coincides with that for Lemos's black string in GR. The existence of $c_1$ and $c_0$ in our expression serves as corrections from the graviton mass to the black string solution and moreover, the structure of horizon will be different. For charged rotating black string the minimum of temperature occurs at $r_+=r_{\text{min}}\equiv \left(\sqrt{c_0+\sqrt{{c_0}^2-(\frac{6\gamma}{\alpha_g\alpha_m})^2}}\right)/{\sqrt{6}}$, while for uncharged rotating black string it appears at $r_+=r_{\text{min}}\equiv\sqrt{c_0/3}$.
The minimum temperature of charged rotating black string reads as
\begin{equation}
T_{\text{min}}=\frac{\left[-72 {\gamma}^2+{\alpha_g}^2\alpha_m^2\Xi\left(6c_0-\sqrt{6}c_1\sqrt{\Xi}\right)\right]\Delta^2}{2\sqrt{6}\pi {\alpha_g}^2\lambda\Xi^{3/2}},
\end{equation}
where $\Xi= c_0+ \sqrt{c_0^2-(6 \gamma/{\alpha_g\alpha_m})^2}$. For uncharged rotating black string, it takes relatively simpler form as follow 
\begin{equation}
T_{\text{min}}=\frac{\alpha_m^2(\sqrt{3 c_0}-c_1)\Delta^2}{2\pi \lambda}, \label{Tmin}
\end{equation} 
which, in the limit $\omega=0, \lambda =1$, reads as follows:
\begin{equation} 
T_{\text{min}}=\frac{\alpha_m^2(\sqrt{3 c_0}-c_1)}{2\pi},
\end{equation}
and matches with the calculated minimum temperature for static black string \cite{Tannukij:2017jtn}.  Let us consider how much the order of magnitude of $T_{\text{min}}$. By reinstalling all constant, Eq. (\ref{Tmin}) can be expressed as
\begin{equation}
\frac{2\pi \lambda}{\Delta^2} T_{\text{min}}=\frac{\hbar \,c}{k_B} \,\alpha_m^2 r_V (\sqrt{3 \bar{c}_0}-\bar{c}_1), \label{Tmin-Dfull}
\end{equation}
where $k_B$ and $\hbar$ are, respectively, Boltzmann constant and Planck constant. Parameters $\bar{c}_0 = c_0/r_V^2$ and $\bar{c}_1 = c_1/r_V$ are dimensionless parameters. As a result, the temperature can be estimated as 
\begin{equation}
T_{\text{min}} \sim T_{\text{Sch}} m_g^2 r_V r_{\text{Sch}} \sim 10^{-27} T_{\text{Sch}}, \label{Tmin-oom}
\end{equation}
where $T_{\text{Sch}}$ is Hawking temperature of Schwarzschild black hole approximated as $10^{-6} \text{K}$ for black hole mass is order of the Sun mass  $M \sim M_S$. The graviton mass $m_g^{-2}$ is approximated as the Hubble radius as $m_g^{-2} \sim H_0^{-2} \sim 10^{46} { \text{km}^2}$ and $r_V \sim 10^{16} \text{km} $. One can see that the minimum temperature is very tiny. Note that one can apply this analysis to the other quantities. For example, we will see later that the remnant size is order of $r_V$.
\begin{figure*}
\begin{tabular}{c c}
\includegraphics[scale=0.64]{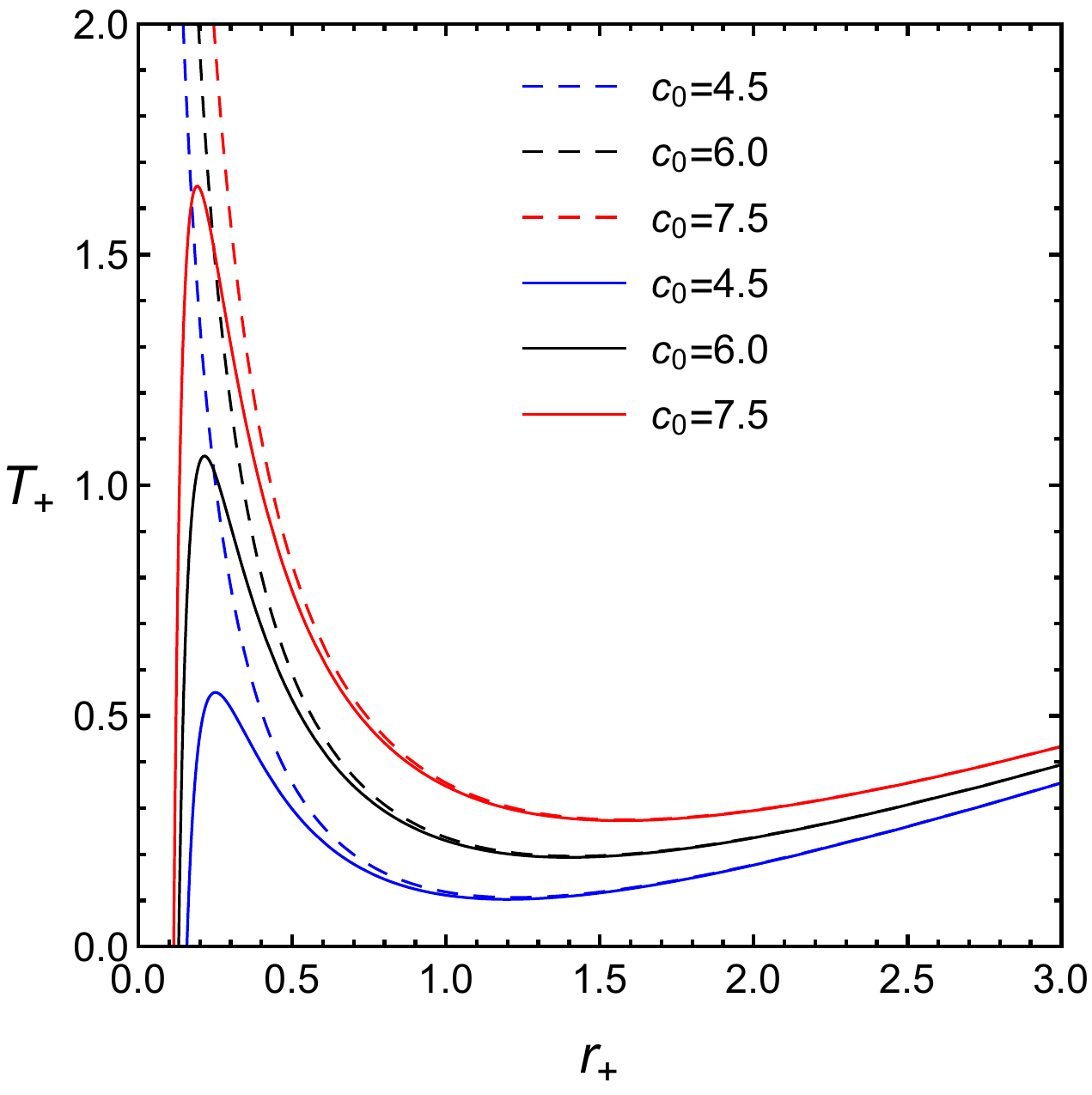}\quad
\includegraphics[scale=0.64]{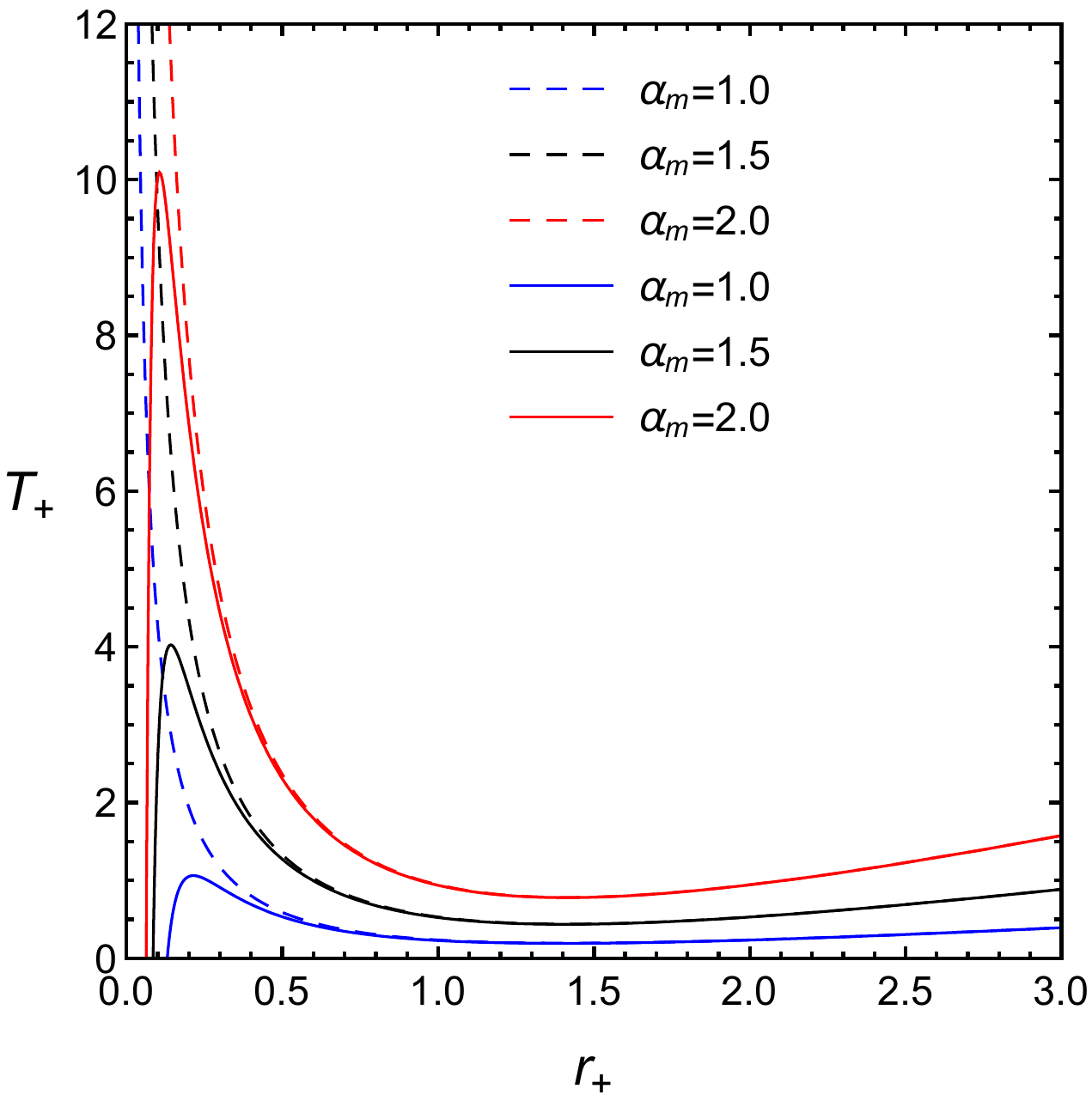}
\end{tabular}
 \caption{The Hawking temperature ($T_+$) vs horizon radius $(r_+)$ for $c_1=3, b=4, \alpha_g=1$. The dashed lines represent the uncharged case, while the solid lines represent charged case with $\gamma = 0.3$. (Left) for varying $c_0$ and (right) for varying $\alpha_m$. (magenta) lines are for Lemos's black strings.}\label{Temp}
\end{figure*}
Horizon temperature of rotating black string is shown in Fig.~\ref{Temp}. During the evaporation process, both uncharged and charged rotating black strings exhibit a local minima in their horizon temperature profile. As the horizon radius further shrink, the horizon temperature increases. Nevertheless, in the last stage of evaporation, charged black string witnesses a zero temperature phase (remnant) where it is cooled down to $T_+=0$ after reaching a finite maximum value, whereas in contrary, the temperature grows monotonically and eventually become unboundedly large for uncharged black string. The remnant size $r_0$ for charged black string reads as 
\begin{eqnarray}
r_0&=&\frac{1}{12}\Big(8c_1^2 -16c_0-4\frac{(P_1+P_3^2)}{P_3\alpha_g^2\alpha_m^2}-\frac{8c_1(c_1^2-3c_0)}{P_4}\Big)^{1/2}\nonumber\\
&&+\frac{1}{6}(c_1-P_4),\label{remnant-size}
\end{eqnarray}
where
\begin{eqnarray}
P_1&=&\alpha_g^2\alpha_m^2(c_0^2\alpha_g^2\alpha_m^2-36\gamma^2),\nonumber\\
P_2&=&2\alpha_g^4\alpha_m^4(c_0^3\alpha_g^2\alpha_m^2+108c_0\gamma^2-54c_1^2\gamma^2),\nonumber\\
P_3&=&\Big(\frac{P_2+\sqrt{P_2^2-4P_1^3}}{2}\Big)^{1/3},\nonumber\\
P_4&=&\sqrt{c_1^2-2c_0+\frac{P_1+P_3^2}{P_3\alpha_g^2\alpha_m^2}}.
\end{eqnarray}
This is clearly inferring from Fig.~\ref{Temp} that graviton mass (in terms of $\alpha_m$) effectively alter the evaporation of black string in dRGT gravity.  Even though remnant size in Eq. (\ref{remnant-size}) is a complicated formula, the order of magnitude of $r_0$ is $c_1 \sim r_V $. While the cut-off scale $\Lambda_3^{-1} \sim 10^{3} \text{km}$ which is much smaller than $r_0  \sim 10^{16} \text{km}$, it implies that the quantum effect may be taken into account at the cut-off scale and then the remnant scale can be justified in classical massive gravity point of view. In particular, as a thermodynamics system, a black hole at zero temperature must  inevitably involve quantum effect. Therefore, more delicate treatment is necessary for addressing this issue. The local maximum value of charged black string temperature increases with increasing $\alpha_m$. The rotating charged black string temperature study reveals that in the late stage of evaporation it will end up to a finite size zero temperature remnant, whose size depends upon various parameters. \newline
Note that the function $\Delta$ in Ref \cite{Lemos:1994xp} is reparameterized to be unity. It is important to note also that without the structure of the graviton mass (for example, setting $c_0 = c_1 =0$), the Lemos's black string horizon temperature does not have extremum points in both charged and uncharged cases (cf. Fig.~\ref{Temp}). Whereas, in the presence of massive graviton, the black string horizon temperature may show both local maxima and minima. Thus, the non-zero graviton mass instigates the second-order thermodynamical phase transitions. This is the crucial result of rotating dRGT black string compared to rotating Lemos's black string; it is possible to obtain the second-order phase transition as well as first-order Hawking-Page transition while it is not for the usual black string.

According to the Kerr-AdS black holes \cite{Gibbons:2004ai}, the first law of thermodynamics and the area law are satisfied by considering the angular velocity measured by the observer relative to a \emph{nonrotating frame} at infinity, $\Omega_{BS}$, instead of one measured by the observer relative to a \emph{rotating frame} at infinity, $\Omega_{H}$. Since dRGT black string solution is asymptotically AdS/dS rather than asymptotically flat, the thermodynamics of the rotating black string should respect actual angular velocity $\Omega_{BS}$. In particular, $\Omega_{BS}$ is defined in a way that the angular velocity of the background spacetime does not contribute. From the black string metric in Eq.~(\ref{bsmetricrotate}) the angular velocity $\Omega$ at $r \rightarrow \infty$ can be found as
\begin{eqnarray}
\Omega_\infty = \lim_{r\rightarrow\infty} \left(-\frac{g_{t\varphi}}{g_{\varphi\varphi}}\right)=\frac{\lambda\omega\left(1-\frac{\alpha_m^2}{\alpha_g^2}\right)}{\lambda^2-\frac{\omega^2\alpha_m^2}{\alpha_g^4}} \label{omegafar}.
\end{eqnarray}
Note that $\Omega_\infty = 0$ when $\alpha_m=\alpha_g$. Thus, we can find $\Omega_{BS}$ as
\begin{eqnarray}
\Omega_{BS}= \Omega_H-\Omega_\infty=\frac{\omega\alpha_m^2\left(\lambda^2-\frac{\omega^2}{\alpha_g^2}\right)}{\lambda\alpha_g^2\left(\lambda^2-\frac{\omega^2\alpha_m^2}{\alpha_g^4}\right)}.
\end{eqnarray}
It is worth mentioning that even if there is no rotating black string present in the spacetime, the background can still possess a nonzero angular momentum due to the difference between $\alpha_m$ and $\alpha_g$ which can be seen in Eq. (\ref{omegafar}), viz., $\Omega_{BS}=\Omega_H$ when $\alpha_m=\alpha_g$. From the one-form of four potential, the electrostatic potential of the black string is $\Phi_q=2q\lambda/(\alpha_g r)$. As a result, with these thermodynamical quantities and the differential form of the first law of thermodynamics
\begin{eqnarray}
TdS = dM - \Omega_{BS}dJ-\Phi_q dq , \label{BH1stlaw}
\end{eqnarray}
we can calculate the entropy per unit length of black string, which yields
\begin{eqnarray}
S_+ &=& \frac{ 1}{2}\pi \lambda \alpha_g r_+^2 = \frac{A}{4},
\end{eqnarray}
where $A=2\pi \lambda \alpha_g r_+^2$ is the black string horizon area per unit length. 
\begin{figure*}[!ht]
\begin{tabular}{c c}
\includegraphics[scale=0.65]{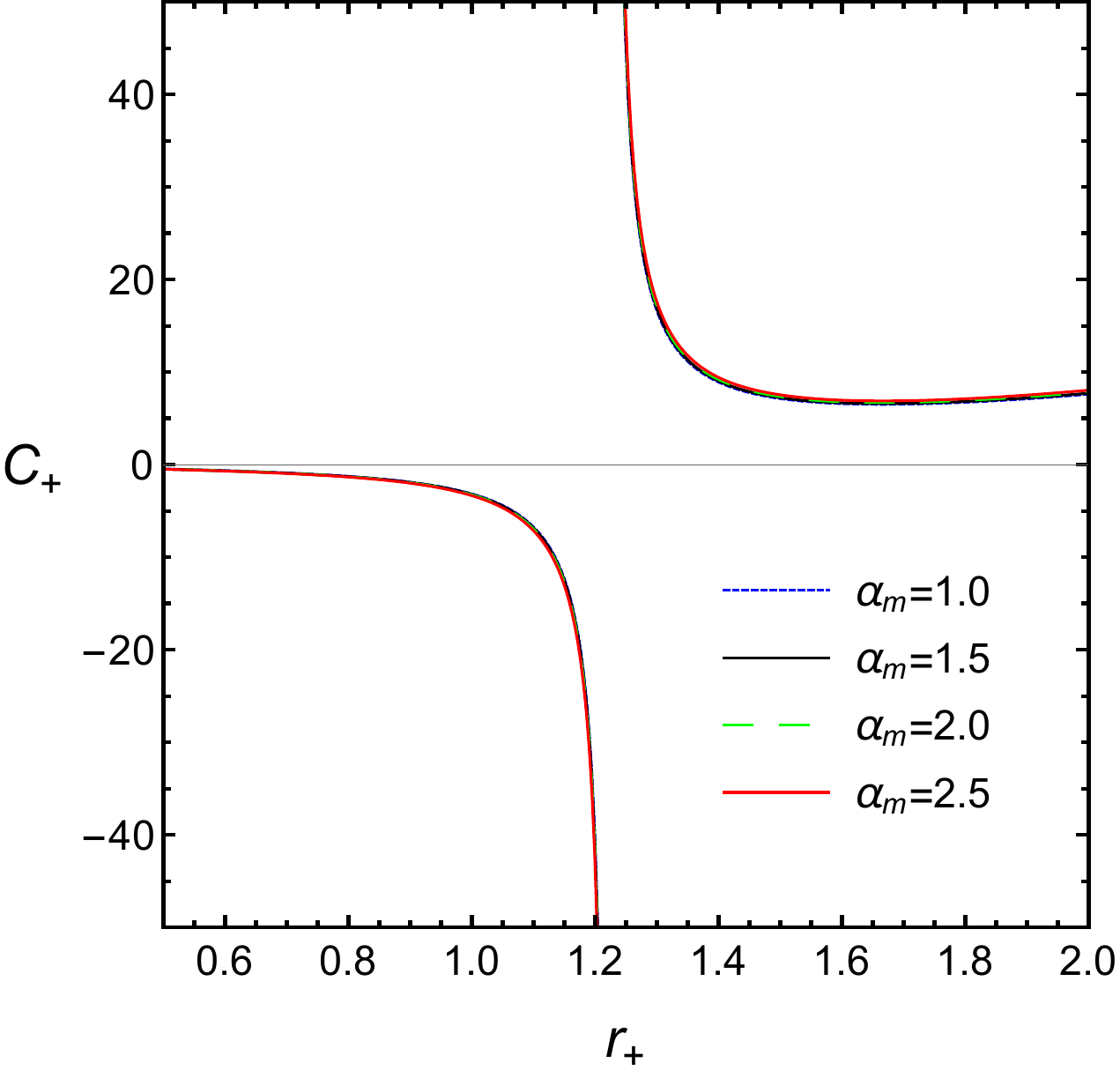}\quad
\includegraphics[scale=0.65]{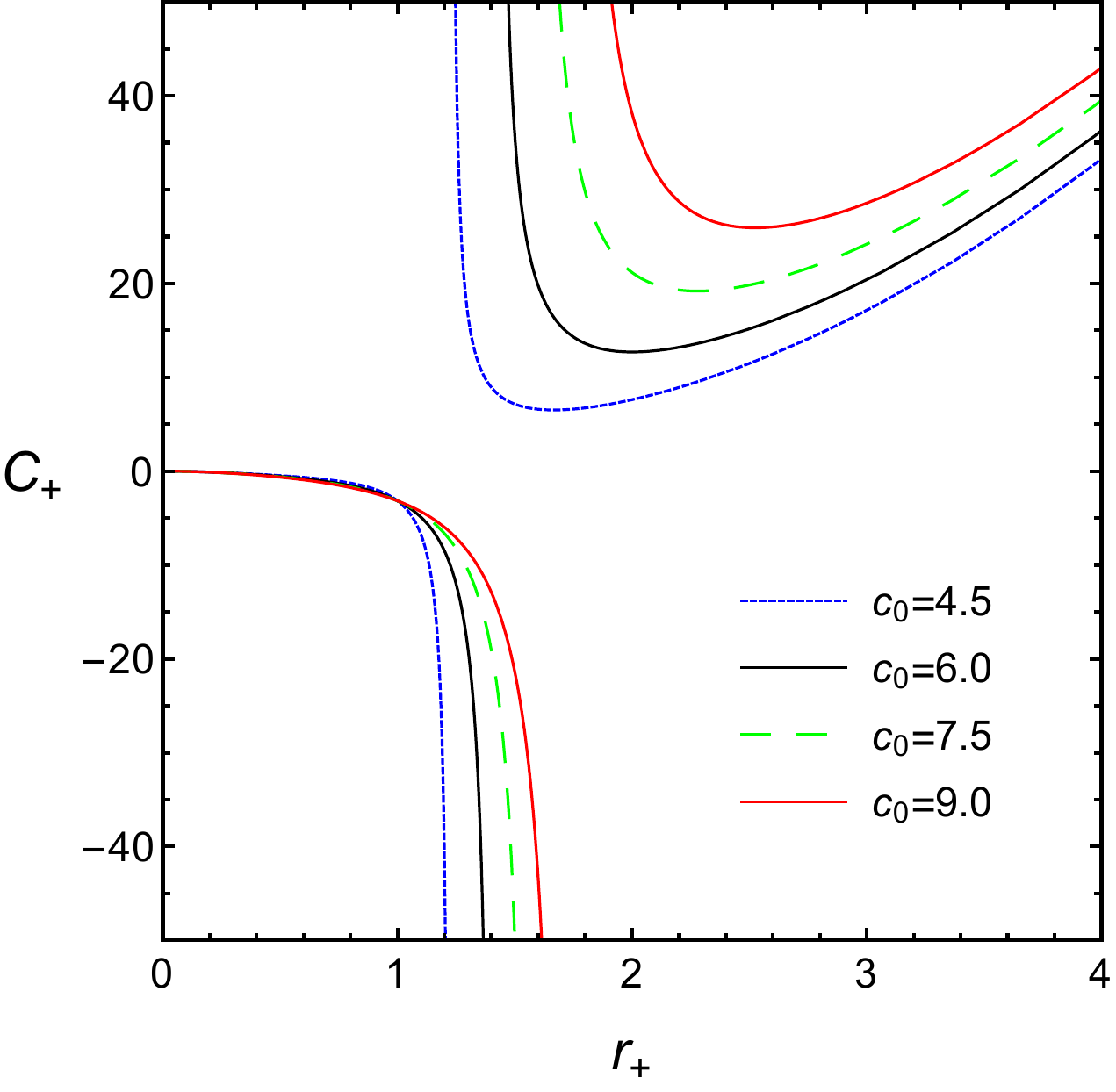}
\end{tabular}
 \caption{The specific heat ($C_+$) behavior with horizon radius ($r_+$) of uncharged rotating black string for parametric values of $\alpha_g=1, c_1=3, \lambda=1,$  and $\omega=0.1$. (Right) for $\alpha_m=1$ and (left) for $c_0=4.5$.  }\label{SpeHeat}
\end{figure*}

In order to determine the black hole local thermodynamical stability, we need to study the specific heat behavior. Local thermodynamical stability signifies that
how the system responds to the small fluctuations in its thermodynamical variables. The heat capacity for charged rotating black string in dRGT massive gravity takes the following form:
\begin{align}
C_+&=\left(T\frac{dS}{dT}\right)_{r=r_+}\nn\\
&=\frac{\pi r_+^2\lambda \alpha_g\left[\gamma^2-r_+^2(3r_+^2-2c_1 r_++c_0)\alpha_g^2\alpha_m^2\right]}{\left[r_+^2(c_0-3r_+^2)\alpha_g^2\alpha_m^2-3\gamma^2 \right]\Delta^2},\,\,
\end{align}
which in the limit $\gamma=0$, gives the value for uncharged rotating black string, reads as
\begin{equation}
C_+=\frac{\pi \alpha_g r_+^2 \lambda\left(3r_+^2 -2c_1 r_+ +c_0\right)}{(3r_+^2-c_0)\Delta^2}.\label{NRSH}
\end{equation}
Furthermore, specific heat for uncharged and static black string can be obtained in the special setting of $\lambda =1, \omega=0, \gamma=0$,
 \begin{equation}
C_+=\frac{\pi \alpha_g r_+^2 \left(3r_+^2 -2c_1 r_+ +c_0\right)}{(3r_+^2-c_0)},
\end{equation}
which matches with the one calculated in Ref.~\cite{Tannukij:2017jtn}. With the specific heat we can analyse the thermodynamical stability as well as the phase transition during Hawking evaporation process. Indeed the local thermodynamical stability of system depends upon the sign of specific heat; if $C_+>0$, then it is stable; otherwise it is unstable.
\begin{figure*}[!ht]
\begin{tabular}{c c}
 \includegraphics[scale=0.65]{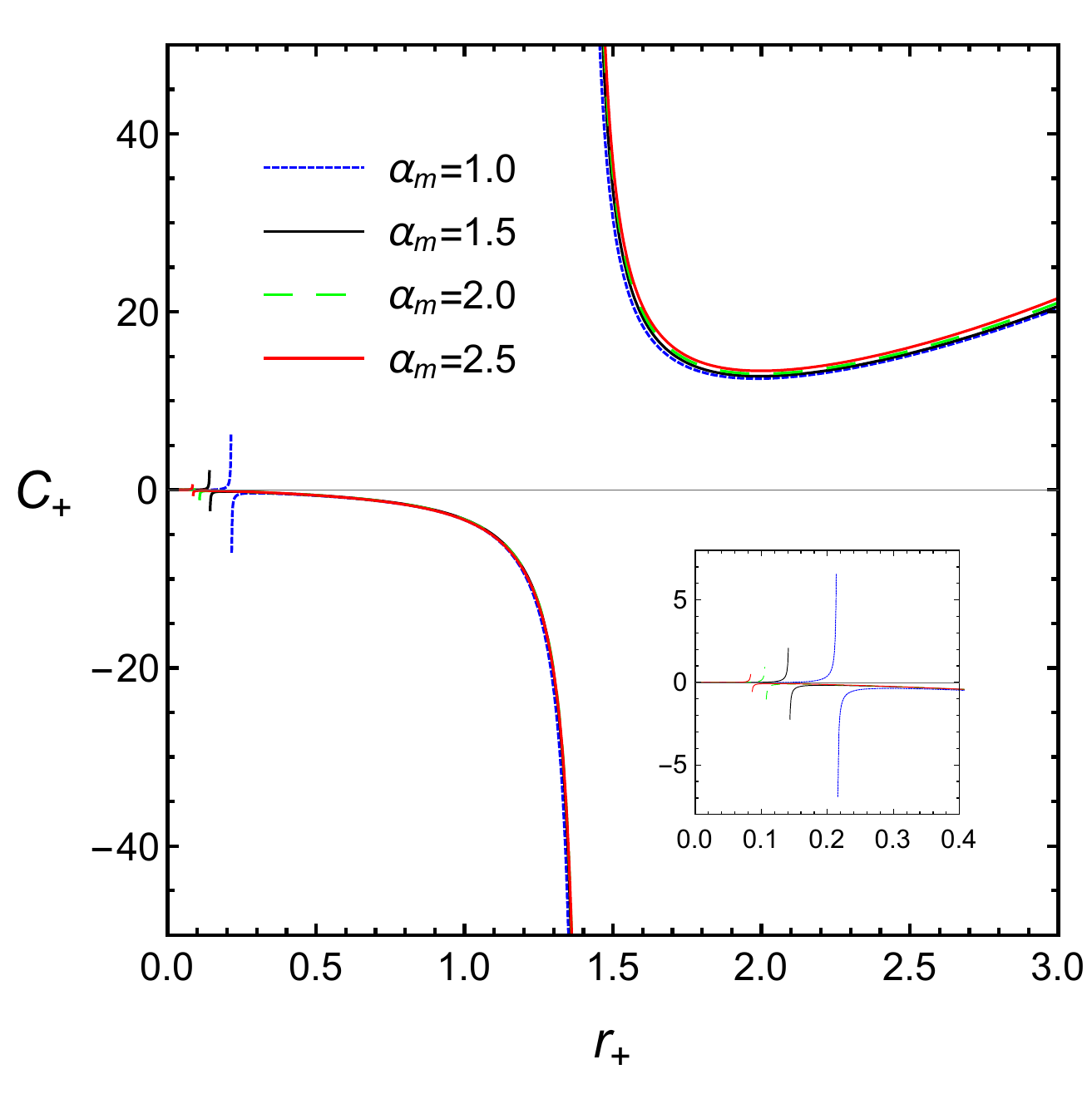}\quad
\includegraphics[scale=0.65]{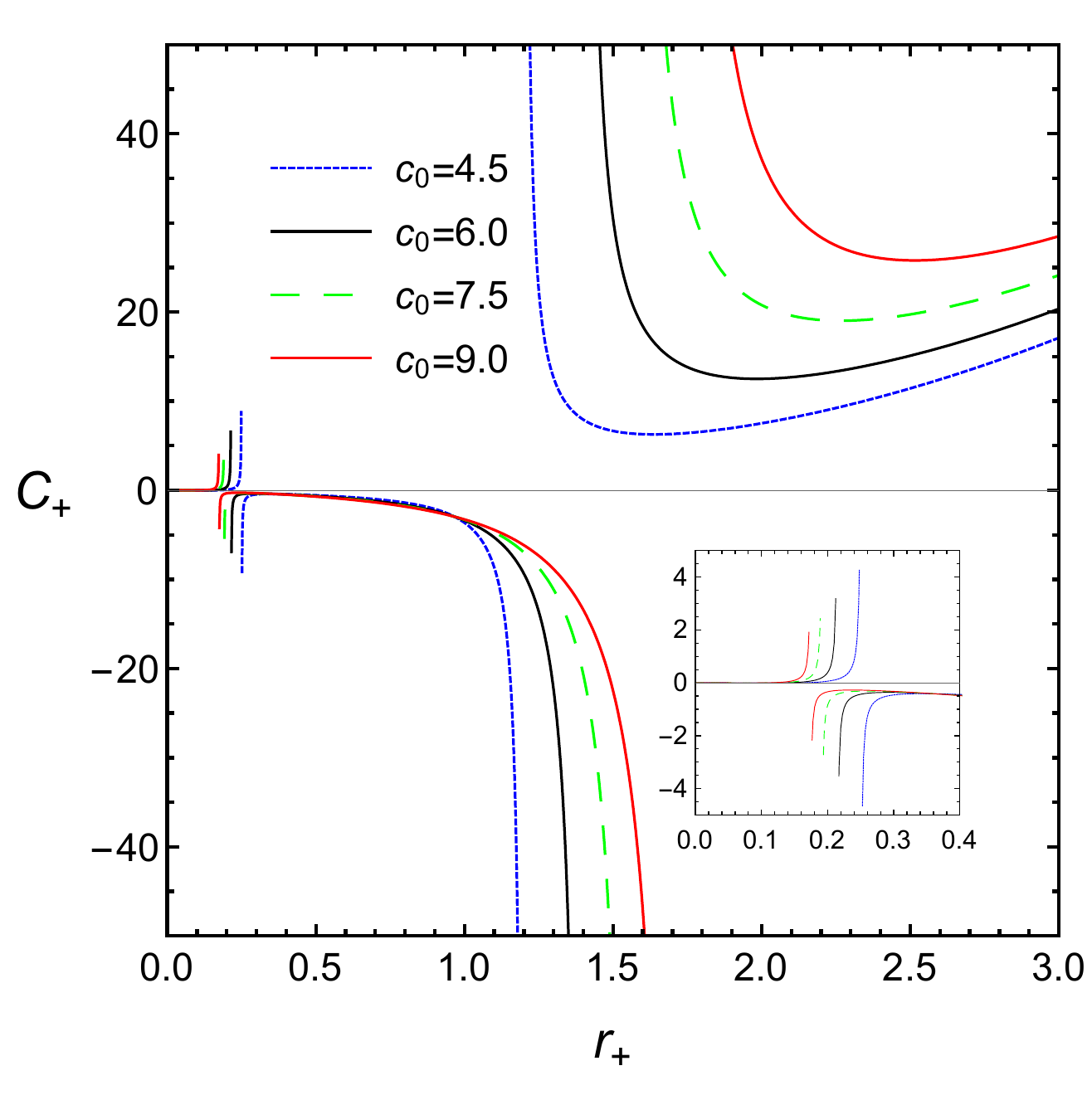}
\end{tabular}
 \caption{Rotating charged black string specific heat ($C_+$) vs horizon radius $(r_+)$ for parametric values of $\alpha_g=1, c_1=3, \lambda=1,$  and $\omega=0.1$. (Right) for $\alpha_m=1$ and (left) for $c_0=6$. }\label{SpeHeat2}
\end{figure*}
Because during the evaporation rotating black string exhibit an extremum temperature, i.e., minima for both uncharged and charged while local maxima only for charged black string, accordingly specific heat $C_+$ will diverge at these extremum points $ r_{\text{min}}$ and $r_{\text{max}}$. As pointed out by Davies \cite{Davies:1978mf}, discontinuity  in specific heat capacity (abruptly changing its sign) implies the second order phase transition in evaporation process. In particular, it is very clear from Eq. (\ref{NRSH}) that for uncharged black string heat capacity is discontinuous at $r_+=r_{\text{min}}\equiv\sqrt{c_0/3}$ and hence it is thermodynamically stable ($C_+>0$) for $r_+>r_{\text{min}}$ while unstable ($C_+<0$) for $r_+<r_{\text{min}}$. It is interesting to find that the phase transition depends upon the structure of graviton mass. In Fig. \ref{SpeHeat}, we have plotted the specific heat ($C_+$) behaviour of uncharged rotating black string with horizon radius $(r_+)$ for varying $\alpha_m$ and $c_0$. As Fig. \ref{Temp} ascertain that the location of minimum temperature remains intact with varying $\alpha_m$, only the temperature value at that point changes. The same situation can be deduced from Fig. \ref{SpeHeat}, the position of second-order phase transition is independent of $\alpha_m$ but significantly vary with changing $c_0$. Black string temperature increases with increasing horizon radius for $r_+>r_{\text{min}}$; hence this is thermodynamical stable in this region, whereas it increases as horizon radius decrease for $r_+<r_{\text{min}}$; consequently, specific heat is negative and it is unstable in this region.

The specific heat of charged black string as a function of horizon radius with varying $\alpha_m$ and $c_0$ is shown in Fig. \ref{SpeHeat2}. Since charged string has a local minima as well as local maxima (cf. Fig. \ref{Temp}) in temperature profile, the specific heat undergoes phase transition at two points $ r_{\text{min}}$ and $r_{\text{max}}$, unlike uncharged string which has only one local minimum in temperature profile and hence exhibits phase transition at only one point. Phase transition at large value corresponds to the $r_{\text{min}}$ whereas that at smaller value is associated with $r_{\text{max}}$.
The position of phase transition (corresponding to the minimum of temperature) changes significantly with varying $c_0$ while it does not change much with $\alpha_m$.
\begin{figure*}[!ht]
\begin{tabular}{c c}
\includegraphics[scale=0.65]{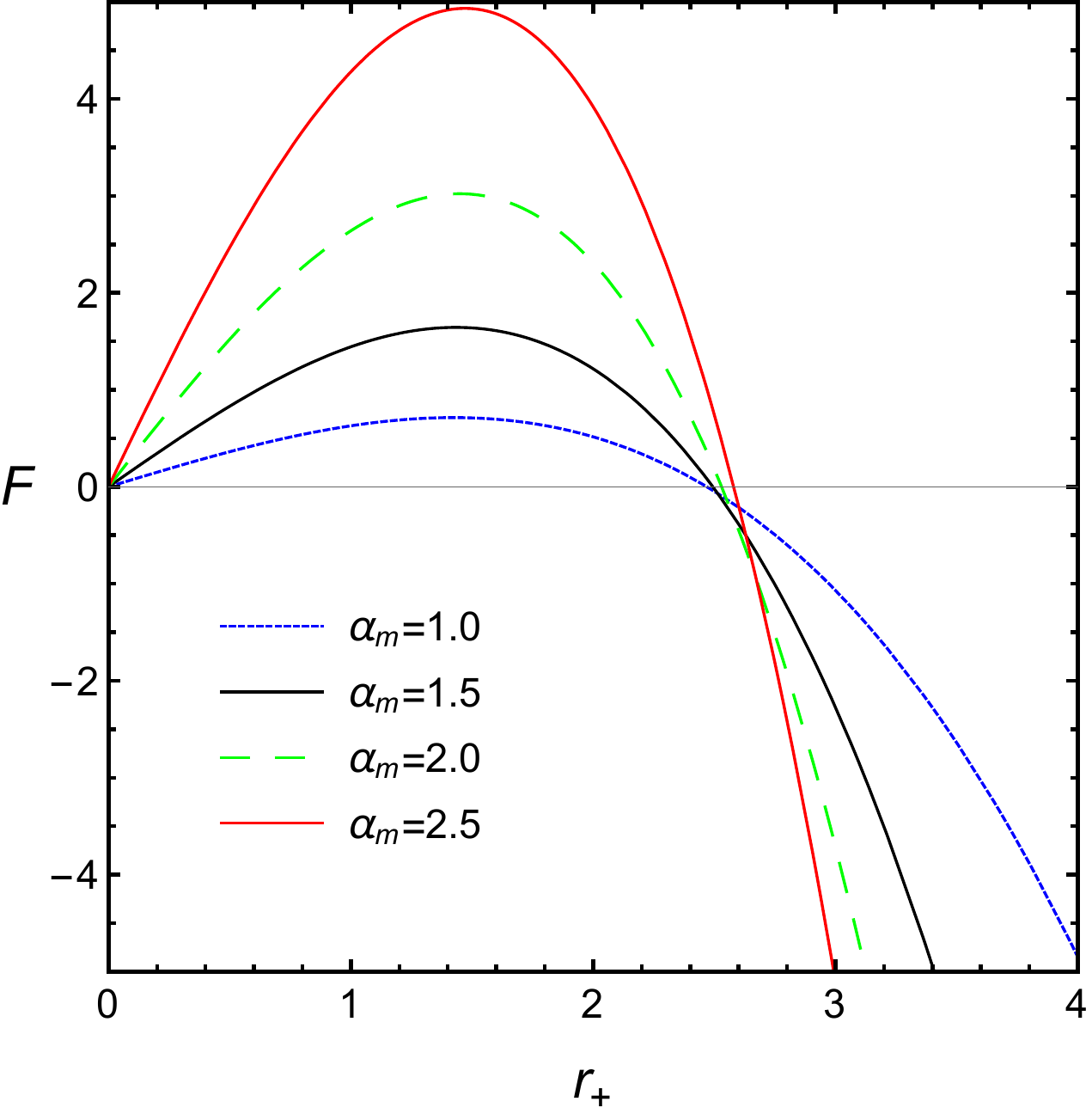}\quad
\includegraphics[scale=0.65]{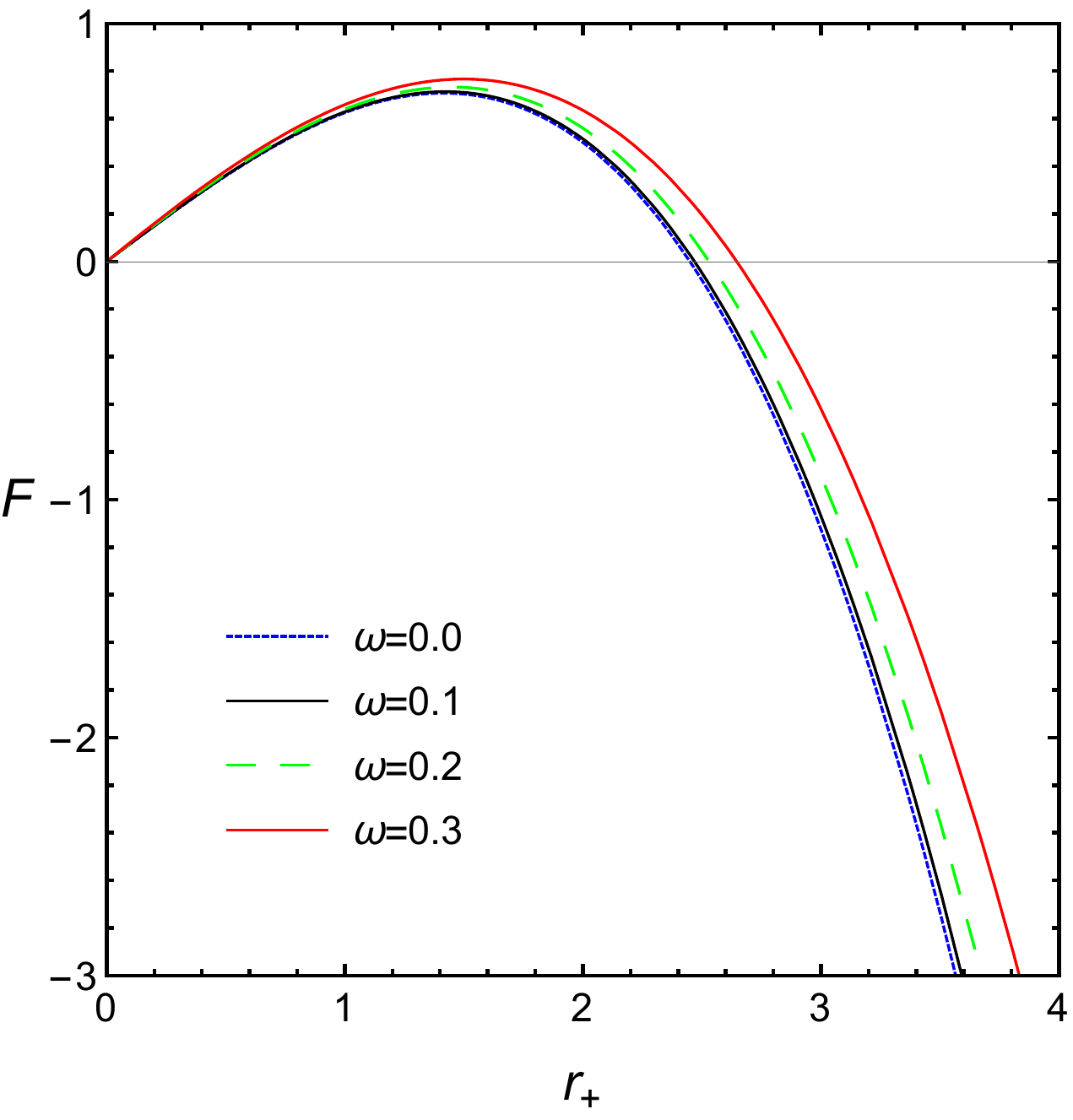}
\end{tabular}
 \caption{The free energy of uncharged black string with various choice of parameters $\alpha_m$ and $\omega$. The other parameters are fixed as follows: $b= 4, \alpha_g = 1, \lambda =1, c_0 =6, c_1 =3$. }\label{F-am-w}
\end{figure*}
The charged rotating black string is thermodynamically stable in the region $r_+>r_{\text{min}}$, whereas thermodynamically unstable in $r_{\text{max}}<r_+<r_{\text{min}}$. Another stable region lies in the interval of zero temperature to the maximum temperature region $r_0\leq r_+\leq r_{\text{max}}$ (cf. Fig. \ref{Temp} and Fig. \ref{SpeHeat2}). It is important to note that without the structure of the graviton mass, $c_0=c_1=0$, the heat capacity of the rotating dRGT black string is positive definite and does not diverge at all, and hence thermodynamical phase transitions are absent. This suggests that the graviton mass may play an important role in high-energy physics of black string where the quantum effects become significant and taken into account. In the presence of multiple horizons associated with spacetime, it is interesting to study its global thermodynamical stability, which is concerned with the phase of a system corresponding to the global
maximum of the total entropy \cite{Hawking:1982dh}.
One can calculate the Helmholtz free energy to discuss the global thermodynamical stability of black string. If we consider that system is in thermodynamical equilibrium with reservoir such that it exchanges only mass $\Delta M\neq 0$ while $\Delta J=\Delta q=0$, then in the preferred phase Helmholtz free energy will be minimum,
\begin{eqnarray}
F&=&M-TS\nonumber\\
&=& \frac{r_+\alpha_g \Delta^2}{8} \left[\frac{\alpha_m^2}{\Delta_{0}^2}\left( (2-3\Delta_{0}^2 ) r_+^2+2 c_1 (\Delta_{0}^2 -1)r_+ \right.\right.\nonumber\\
&&\left.\left.\quad- (\Delta_{0}^2 -2)c_0\right)+\frac{\gamma^2(\Delta_{0}^2 +2)}{r_+^2\alpha_g^2 \,\,\Delta_{0}^2} \right].\label{free-energy}
\end{eqnarray}

In order to find the global stability of the black string, one has to find the condition where the free energy is negative.  Therefore, one can solve $F=0$ for $r_+$. It is convenient to firstly consider the uncharged case. As a result, the critical horizon radius can be expressed as 
\begin{eqnarray}
r_c &=&\frac{c_1 (\Delta_{0}^2-1)}{3 \Delta_{0}^2 - 2} \left[1-\sqrt{1-\frac{(3\Delta_{0}^2-2)(\Delta_{0}^2-2)}{(\Delta_{0}^2-1)^2 c_1^2/c_0}}\right].\nonumber\\
\end{eqnarray}
As a result, one found that the structure of the graviton mass significantly provides the first-order Hawking-Page phase transition between globally stable black string and AdS-like background of massive gravitons. 
From the condition of positive definite of the thermodynamical mass of the black string, one can restrict our attention to the case of small angular frequency, $\omega \ll \lambda\alpha_g^2/\alpha_m$. Then one obtains the approximation as 
\begin{eqnarray}
r_c \approx \sqrt{c_0} + (2\sqrt{c_0} - c_1)\frac{\omega^2 \alpha_m^2}{\alpha_g^4}.\label{critical horizon}
\end{eqnarray}

\begin{figure*}[ht]
	\begin{tabular}{c c}
		\includegraphics[scale=0.65]{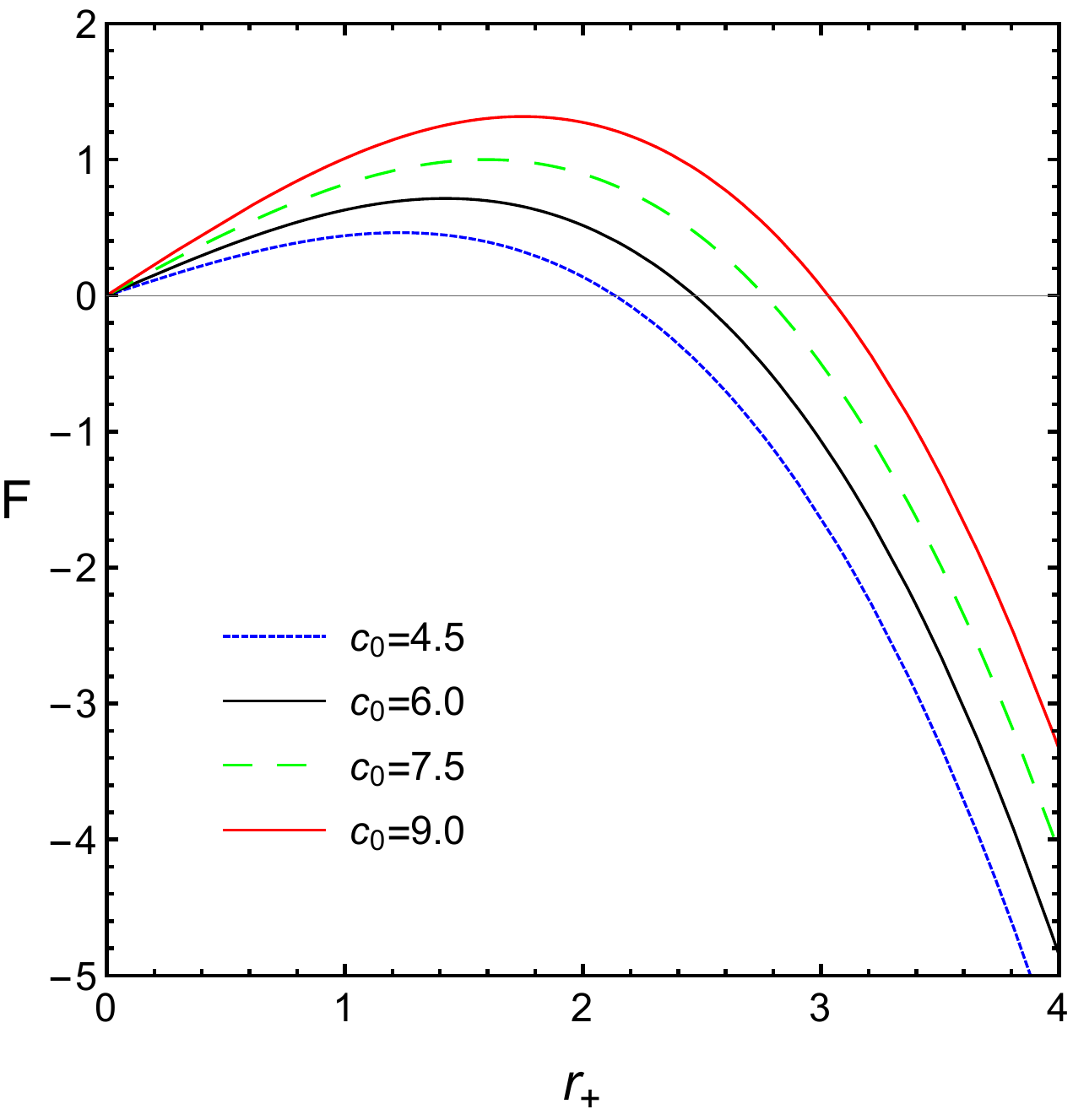}\quad
		\includegraphics[scale=0.65]{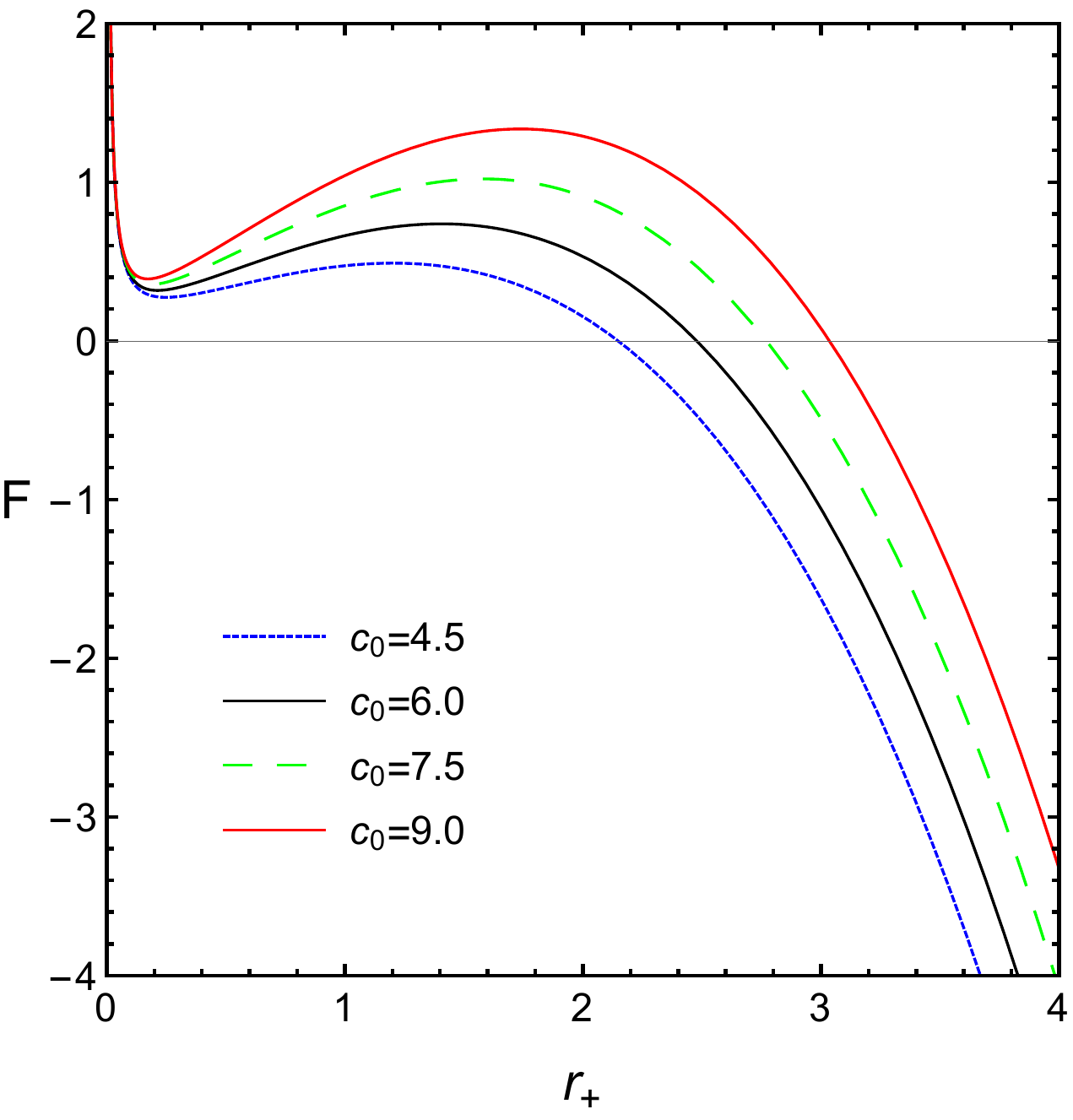}
	\end{tabular}
	\caption{Rotating uncharged (left) and charged (right) black string free energy ($F$) behaviour with horizon radius ($r_+$). }\label{FreeEn}
\end{figure*}

Note that the other value is always negative and we do not consider that here. From this critical horizon, one can see that it reduces to the non-rotating case, $r_c = \sqrt{c_0}$ for $\omega =0$ as found in  \cite{Tannukij:2017jtn}. Moreover, if we impose the existence of the minimum potential, $c_0 > c_1^2 /3$, it is found that the critical horizon is always larger than that from the non-rotating case. As a result, the rotating black string is thermodynamically stable with the horizon larger than its nonrotating counterparts. These results can be confirmed by using numerical plot as shown in Fig. \ref{F-am-w}. Since the value of critical horizon radius is characterized by two parameters, graviton mass parameter $\alpha_m$ and angular frequency $\omega$, the figures show that the more value of the parameters, the larger the critical horizon as indicated from Eq. (\ref{critical horizon}). This result is also true for the charged case, but the equation is very lengthy and is not provided here. The effect of parameter $c_0$ on the free energy is also shown explicitly in Fig. \ref{FreeEn} for both uncharged and charged cases. As we can infer from Eq. (\ref{critical horizon}), the more value of $c_0$ leads to the bigger stable black string. It is important to note that there is a tiny range of parameters to provides the free energy to be positive again for the larger horizon. From  Eq. \eqref{free-energy}, this can occur when the angular frequency $\omega$ is large enough, for example, $\Delta_{0}^2 <2/3$ and $(3\Delta_{0}^2-2)(\Delta_{0}^2-2)c_0 < c_1^2 (\Delta_{0}^2-1)^2$. This case will not be considered here since it may encounter the fine tuning in parameters, and we aim to find how the thermodynamics properties of the black string change when the black string gets small rotation. Note also that the la\fixme{r}ger value of $\omega$ will violate the positive definite condition of the thermodynamics mass as we have mentioned earlier.
\section{Concluding remarks}\label{summary}
In this paper, we obtained both uncharged and charged rotating black string solutions in dRGT massive gravity theory. All the limiting cases are discussed. The obtained solutions are naturally AdS/dS-like depending upon the suitable choices of the parameters, which naturally stem from the non-zero graviton mass in the dRGT gravity theory. Since the black string has an infinite extension along the $z-$axes, therefore it is worthy to determine the actual physical quantities associated with it.We use the Hamiltonian formalism to extract the mass density, angular momentum density, and charge density. In the thermodynamical analysis, we found that the nonzero graviton mass significantly alters the horizon temperature profile during the evaporation process. The first law of black hole mechanics still holds valid for rotating black string.We have also analyzed the thermodynamical stability of both charged and uncharged rotating black strings. It is found that rotating black string in dRGT massive gravity theory, both uncharged and charged, undergoes a second-order phase transition during the Hawking evaporation process, contrary to the Lemos's black string. Furthermore, charged black string also leads to a zero-temperature remnant of finite size in the last stage of evaporation. We must emphasize that, in the semiclassical description of thermodynamics, we cannot precisely determine the fate in the last stage of evaporation as the quantum effects will become dominate. Therefore, this issue requires a more delicate treatment; nevertheless, the presented study is just a first step in the analysis. For global thermodynamical stability, we also studied the behavior of free energy and it is found that the stable rotating black string is bigger than the nonrotating one.   

\section*{Acknowledgement}
This project is supported by the ICTP through Grant No. OEA-NT-01. S. G. G. would like to thank DST INDO-SA bilateral project DST/INT/South Africa/P-06/2016, and
SERB-DST for the ASEAN project IMRC/AISTDF/CRD/2018/000042. R. K. thanks UGC, Government of India for financial support through SRF scheme. P.W. is supported by the Thailand Research Fund through Grant No. MRG6180003. L. T. is supported by the National Research Foundation of Korea grant funded by the Korea government (MSIP) (Grant
No. 2016R1C1B1010107). The authors acknowledge King Mongkut's University of Technology Thonburi for funding Postdoctoral Fellowship to L. T.


\end{document}